\documentclass[fleqn,useAMS]{Papers}
\usepackage{newtxtext,newtxmath}
\usepackage[T1]{fontenc}
\usepackage{subfloat}
\usepackage{makecell}
\usepackage[square,sort,numbers]{natbib}
\usepackage{graphicx}
\usepackage{amsmath}
\usepackage{lipsum}
\usepackage{caption}
\usepackage{comment}
\usepackage{lineno}
\usepackage{xcolor}
\usepackage[normalem]{ulem}
\newcommand{\added}[1]{{\color{black}#1}}

\title[From turbulence to cosmic structure]{From turbulence to cosmic structure: a unified stochastic random walk theory for Kolmogorov spectrum, halo mass functions, and density profiles}

\author[Z. Xu]{Zhijie (Jay) Xu,$^{1}$\thanks{E-mail: \href{mailto:zhijie.xu@pnnl.gov}{zhijie.xu@pnnl.gov}; \href{mailto:zhijiexu@hotmail.com}{zhijiexu@hotmail.com}}
\\
$^{1}$Physical and Computational Sciences Directorate, Pacific Northwest National Laboratory; Richland, WA 99354, USA\\
}

\date{Accepted XXX. Received YYY; in original form ZZZ}
\pubyear{2025}

\begin{document}
\label{firstpage}
\pagerange{\pageref{firstpage}--\pageref{lastpage}}
\maketitle

\begin{abstract}
We develop a unified stochastic framework for a broad class of nonequilibrium systems and apply it to turbulence and cosmic structure formation. The evolution of spectra and distributions in these systems are modeled as transport processes in an appropriate "scale space", described by random walks and Fokker-Planck equations with scale-dependent waiting times. We reformulate the turbulent cascade as a random walk in wavenumber space with a k-dependent waiting time $\tau_{gk}\propto k^{-{\gamma}}$. With self-consistency condition determined value $\gamma=2/3$, we recover the Kolmogorov -5/3 law and obtain the full energy spectrum including the inertial-dissipation crossover, exponential cutoff, and spectrum constant without invoking dimensional analysis directly on the energy spectrum. For a bottom-up hierarchical halo structure formation, low-mass haloes collapse early and retain strong primordial imprints, while prolonged assembly of large haloes near the characteristic mass $m_h^*$ tends to erase this memory and establishes a universal spectrum $P(k)\propto k^{-1}$. We model this process as an analogous random walk of dark matter haloes in mass space $m_h$ with a mass-dependent waiting time $\tau_g\propto m_h^{-\lambda}$. The model yields a stretched Gaussian halo mass function that reproduces the Press-Schechter form, with both small-scale scaling $f_M\propto m_h^{-\lambda}$ and exponential cutoff at $m_h>m_h^*$ set by exponent $\lambda$. The framework naturally produces a double-$\lambda$ mass function, where $\lambda$ reflects the primordial spectrum at low masses and approaches $\lambda$=2/3 near $m_h^*$ where gravity dominates. For individual haloes, a spatial random walk with waiting time $\tau_{gr}\propto r^{-\gamma}$ generates a double-$\gamma$ density profile that reproduces the Einasto profile, with both small-scale scaling $\rho_r\propto r^{-2\gamma}$ and exponential cutoff at $r>r_s$ set by exponent $\gamma$. In this case, $\gamma$ is fixed by the initial spectrum in the inner core regions and approaches $\gamma$=2/3 and $\rho_r\propto r^{-{4/3}}$ in the outer regions when gravity dominates. Our results demonstrate that this simple stochastic framework offers a common theoretical description of both turbulent cascades and cosmic structure formation.
\end{abstract}

\begin{keywords}
\vspace*{-10pt}
Turbulence; Energy spectrum; Dark matter; Mass function; Halo density;
\end{keywords}

\begingroup
\let\clearpage\relax
\tableofcontents
\endgroup
\vspace*{-25pt}

\section{Introduction}
\label{sec:1}
Turbulence plays a central role in modern astrophysics. Most astrophysical plasmas and gases are neither laminar nor static. From the interstellar medium (ISM) to galaxy clusters, accretion disks around compact objects, and stars, turbulent flows are ubiquitous, typically highly compressible, magnetized, and driven far from equilibrium \citep{Elmegreen:2004-Interstellar-Turbulence,Falceta-Gonçalves:2014-The-onset-of-large-scale-turbulence,Federrath:2012-THE-STAR-FORMATION-RATE-OF-TURBULENT}. In such environments, the detailed microphysics is often poorly constrained, but the large-scale dynamics are dominated by cascades of energy across many orders of magnitude in scale. Turbulence theory provides the natural framework for describing this multiscale transfer of energy, momentum, and magnetic helicity.

Because microphysics is usually uncertain or inaccessible, astrophysics relies heavily on the idea of universal turbulent spectra. Kolmogorov's -5/3 law and its \added{magnetohydrodynamic (MHD)} extensions provide key relations between the mean energy flux and the distribution of energy in wavenumber (k) space. If the spectrum is universal and dependent only on a few macroscopic parameters, one can extrapolate from laboratory experiments and numerical simulations to regimes far beyond terrestrial conditions. These considerations motivate the search for a universal spectrum not only for ordinary baryonic turbulence but also for more exotic sectors. If dark matter, for example, undergoes some form of mass/energy transfer through gravitational clustering, one may ask whether its distribution and structure tend toward a universal form analogous to the Kolmogorov spectrum, manifested in halo mass functions and density profiles \citep{Xu:2023-Dark-matter-halo-mass-functions-and, Xu:2021-Inverse-mass-cascade-mass-function}. Establishing such universality would provide a powerful bridge between microphysical dark matter models and macroscopic cosmological observables \citep{Xu:2022-Postulating-dark-matter-partic}.

Nonlinear nonequilibrium systems in both fluid dynamics and cosmology exhibit a remarkably similar phenomenology: strongly self-interacting many-body dynamics give rise to robust, nearly universal spectra and distributions when viewed in an appropriate ``scale space''. Turbulent flows provide the canonical example. At high Reynolds number, energy injected at large scales $L$ cascades through a hierarchy of eddies toward small scales, where it is dissipated at the dissipation scale $\eta$. In the classical Kolmogorov (K41) framework, intermediate inertial-range scales are assumed to be statistically homogeneous, isotropic, and self-similar, and to be governed solely by the mean energy flux $\varepsilon$ through scales. Under these assumptions, the dimensional analysis leads to the celebrated Kolmogorov -5/3 spectrum \citep{Kolmogoroff:1941-Dissipation-of-energy-in-the-l}. Outside the inertial range, viscous effects determine the spectral shape, and empirical ``bridge'' functions are commonly employed to smoothly interpolate between the inertial- and dissipation-range behavior. A widely used empirical spectrum often takes the form (Pao spectrum \citep{Pao:1965-Structure-of-Turbulent-Velocity})
\begin{equation} 
\label{eq:1-1} 
E(k)=C_K \varepsilon^{2/3}k^{-5/3}\exp\left[-\frac{3}{2}C_K(\eta k)^{4/3}\right],
\end{equation}
where $\eta$ is the Kolmogorov microscale and $C_K$ is a universal constant of value 1.5 to 1.7. Thus, despite the complexity of the Navier-Stokes dynamics, the energy spectrum assumes a simple, nearly universal form: a -5/3 power law with an exponential cutoff at large $k$, set primarily by the mean dissipation rate $\varepsilon$ and only weakly dependent on the details of the initial forcing. Nevertheless, deriving such a universal spectrum from first principles, using only exact conservation and transfer relations, and identifying the spectrum constant $C_K$ in terms of physical parameters, remains an outstanding theoretical problem and provides the primary motivation for this work.
  
Remarkably similar issues also arise in the nonlinear evolution and flow of collisionless dark matter \citep{Xu:2023-Universal-scaling-laws-and-density-slope,Xu:2023-On-the-statistical-theory-of-self-gravitating,Xu:2024-On-the-statistical-theory-of-self-gravitating-high-order}. Gravitational collapse produces highly localized, virialized haloes that are the fundamental building blocks of large-scale structures \citep{Cooray:2002-Halo-models-of-large-scale-str}. Their number density as a function of mass is encoded in the halo mass function, while their internal structure is described by halo density profiles. The Press-Schechter approach and its generalizations model halo formation as a first-crossing problem, in which a stochastic trajectory in smoothed density space crosses a critical collapse barrier \citep{Press:1974-Formation-of-Galaxies-and-Clus,Bond:1991-Excursion-Set-Mass-Functions-f}. Subsequent extensions to correlated random walks further refined this framework \citep{Musso:2012-One-step-beyond-the-excursion-set-approach,Paranjape:2012-Peaks-theory-and-the-excursion-set-approach,Maggiore:2010-THE-HALO-MASS-FUNCTION-FROM-EXCURSION-SET-THEORY}. The resulting Press-Schechter (PS) mass function closely parallels the turbulence energy spectrum $E(k)$ in Eq.~\eqref{eq:1-1}, exhibiting a power-law behavior at low masses, controlled by the effective spectral index $n$, and an exponential cutoff $M_{PS}^*$,
\begin{equation}
f_{PS} \left(m_{h} \right)=\frac{3+n}{3\sqrt{2\pi}}\frac{1}{{M_{PS}^*}}\left(\frac{m_h}{{M_{PS}^{*}}}\right)^{-\frac{3-n}{6} }\exp \left[-\frac{1}{2} \left(\frac{m_{h}}{{M_{PS}^{*}}}\right)^{\frac{3+n}{3}} \right],
\label{eq:1-2}
\end{equation}
where $m_h$ is the mass of haloes. Despite providing a remarkably simple analytic estimate of halo abundances, the PS mass function relies on highly idealized assumptions, most notably spherical/ellipsoidal collapse (\citep{Press:1974-Formation-of-Galaxies-and-Clus,Sheth:2001-Ellipsoidal-collapse-and-an-im}) and a fixed critical overdensity $\delta_c$, thereby motivating alternative perspectives in this paper.

The halo density profile from highly nonlinear gravitational dynamics poses another problem closely analogous to the appearance of universal spectrum in turbulence. \added{The widely used NFW profile \citep{Navarro:1997-A-universal-density-profile-fr} instead has fixed inner ($-1$) and outer ($-3$) slopes with no exponential cutoff, and thus does not share this structural correspondence; it remains an excellent empirical fit over the radial range typically resolved in simulations, though with one fewer free parameter than Einasto, whose additional shape parameter $\alpha$ is expected to improve fit quality.} Over a wide range of halo masses and cosmological models, numerical simulations show that the spherically averaged density profiles are well described by the Einasto form, which \added{like the energy spectrum (Eq.~\eqref{eq:1-1}) and halo mass function (Eq. \eqref{eq:1-2}),} features an exponential cutoff \citep{Navarro:2004-The-inner-structure-of-ACDM-ha, Navarro:2010-The-diversity-and-similarity-of-simulated, Ludlow:2013-The-mass-profile-and-accretion, Ludlow:2016-The-mass-concentration-redshif}
\begin{equation}
\label{eq:1-3} 
\rho_{h}^{ES}(r)= \rho_s \exp\left(\frac{2}{\alpha}\right) \exp\left[-\frac{2}{\alpha}{\left(\frac{r}{r_s}\right)^{\alpha}}\right],
\end{equation}
where $\rho_s$ is the density at scale radius $r_s$. The shape parameter $\alpha$ varies weakly with mass and redshift, indicating a robust, approximately self-similar structure across haloes of different scales \citep{Gao:2008-The-redshift-dependence-of-the}. Despite intensive study, there is currently no first-principle theory that uniquely predicts the density profile and the physical meaning of $\alpha$, which also motivates the present work.

These parallels motivate a unified viewpoint in which three seemingly distinct problems (Eqs.~\eqref{eq:1-1}, \eqref{eq:1-2}, and \eqref{eq:1-3}) can be treated within a common framework. Because the dark component is collisionless, it does not support turbulent cascades driven by viscous dissipation and vortex; similarly, a halo that has reached full virial equilibrium cannot sustain a net transfer of mass or energy across spatial scales. Consequently, we acknowledge the difference between three problems and focus on the scale-to-scale redistribution of mass and energy in self-gravitating dark matter, rather than a viscous Kolmogorov-type cascade for dark matter; the "cascade" of Section~\ref{sec:3} refers to this redistribution in halo mass space, carried by mergers and accretion among haloes, not to a viscous cascade.

\begin{figure}
\includegraphics*[width=\columnwidth]{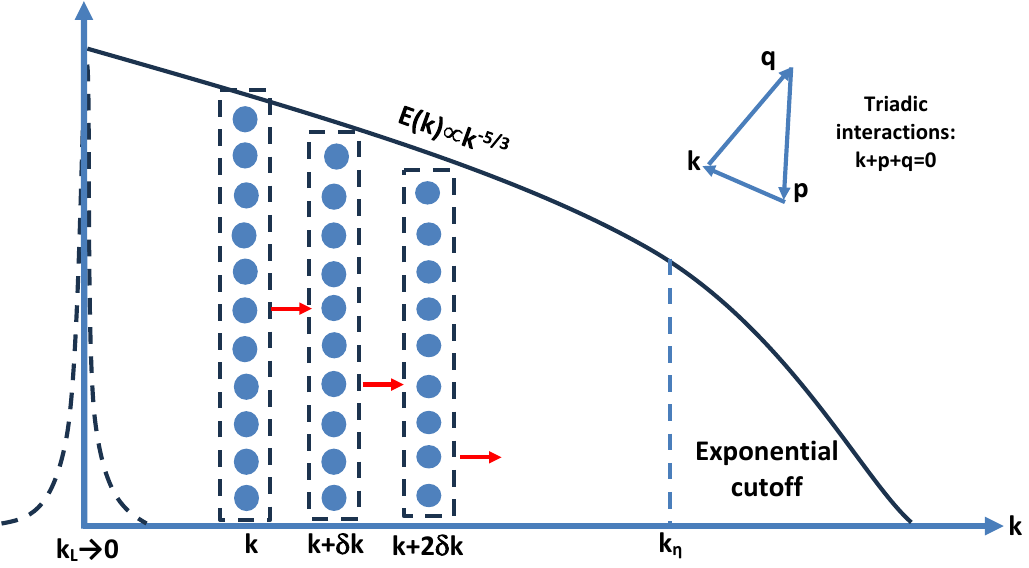}
\caption{Schematic depiction of a random walk in wavenumber k space underlying the turbulence energy spectrum. Energy is injected at the integral scale $k_L$ (or equivalently $L$ in real space), represented by a delta-like distribution of a group of ``energy carriers'', each carrying an infinitesimal energy $dE$. The dominant local triadic interactions ($\textbf{k+p+q}$=0) transfer "energy carriers" (blue dots) from a wavenumber k to a nearby wavenumber k+$\delta$k, producing a step-wise random walk toward larger k (red arrows). Each step conserves total energy, while the waiting time $\tau_{gk}$(k) between steps depends on k and encodes the finite lifetime of eddies at that scale. In the stationary state, the resulting distribution $P(k)$ of "energy carriers" is proportional to the energy spectrum $E(k)$, shown here as a Kolmogorov -5/3 law with an exponential cutoff near the dissipation wavenumber $k_{\eta}$ (or $\eta$ in real space), where viscous effects dominate and remove energy.} 
\label{fig:1}
\end{figure}

The unified framework employs random-walk or diffusion models in an appropriate ``scale space''. In this view, the ``cascade'' or ``scale-to-scale'' mass or energy transfer is formulated as a stochastic transport process in that scale space. Using the turbulence energy spectrum as an example (Fig. \ref{fig:1}), the ``cascade'' is driven by local triad interactions and constrained by exact energy conservation. The energy spectrum is interpreted as the distribution of ``energy carriers'' carrying an infinitesimal energy $dE$ and executing a random walk in k-space, with a scale-dependent waiting time $\tau_{gk}(k)$ that reflects the nonlinear eddy turnaround time. The stationary solution of the corresponding Fokker-Planck equation in k-space, with viscous dissipation acting as a scale-dependent sink (Eq.~\eqref{eq:2-5}), yields the entire energy spectrum. This simple framework recovers the Kolmogorov -5/3 law, predicts the exponential dissipation-range cutoff, and fixes the spectrum constant $C_K$. The k-independence of the mean energy flux $\varepsilon$ in the inertial range is not assumed a priori but emerges as a consequence of the steady-state Fokker-Planck equation. 

In what follows, we apply this random-walk framework to three seemingly distinct problems: (i) the turbulence energy cascade as a random walk in wavenumber space $k$, yielding the Kolmogorov -5/3 scaling and the entire energy spectrum; (ii) hierarchical halo assembly as a random walk in halo mass space $m_h$, yielding the halo mass function; and (iii) particle migration within haloes as a random walk in space $\textbf{x}$, shaping the halo density profile. In each case, the many-body nonlinear dynamics are coarse-grained into a continuous-time random walk in an appropriate "scale space" $x$. 
The walk reduces to a genuine Fokker-Planck equation subject to two conditions: 1) spatial scale separation, requiring the local jump size to be small compared to the scale over which the waiting time $\tau(x)$ itself varies, so that a position-dependent diffusivity $D(x)$ is locally well defined; 2) temporal scale separation, requiring many elementary transfer events to occur within each macroscopic evolution time, so that the mean waiting time at each scale, instead of a full distribution of waiting time, controls the coarse-grained dynamics \citep{Xu:2021-Inverse-mass-cascade-mass-function}. Universal spectra or distributions naturally emerge as stationary or time-dependent solutions. We develop this perspective systematically, constructing random-walk models in each "scale space", and demonstrate how this unified framework can account for the observed universality of turbulence spectra, halo mass function, and density profiles. 

\section{Random walk in k-space for turbulence}
\label{sec:2}
Kolmogorov's -5/3 scaling is derived under the assumption that, at sufficiently high Reynolds numbers, small-scale turbulence is statistically homogeneous and isotropic and is characterized solely by a scale-independent mean energy dissipation rate $\varepsilon$ and the kinematic viscosity $\nu$. In the inertial range, between $k_L$ and $k_{\eta}$ (Fig.~\ref{fig:1}), energy is transferred conservatively across scales. Dimensional analysis then implies a universal energy spectrum $E(k)=C_K\varepsilon^{2/3}k^{-5/3}$ over this range. In the following, we show that the energy cascade can instead be formulated as a random walk process in wavenumber space, from which the -5/3 scaling, the full spectral shape, and the spectrum constant $C_K$ emerge naturally without requiring a scale-independent dissipation rate $\varepsilon$ as a separate assumption.

We first normalize the steady state energy spectrum to obtain a probability density $P(k)$ in wavenumber space. The total kinetic energy per unit mass satisfies
\begin{equation} 
\label{eq:2-1} 
\int_0^{\infty} E(k)dk=\frac{1}{2}u_0^2,
\end{equation} 
where $u_0$ is the characteristic velocity at the integral scale $L$. We then define the distribution function $P(k)$
\begin{equation} 
\label{eq:2-2} 
\begin{split}
&P(k) = 2{E(k)}/{u_0^2} \quad \textrm{such that} \quad \int_0^{\infty} P(k) dk=1, \\
&N(k) = P(k)k=2{E(k)k}/{u_0^2}.
\end{split}
\end{equation} 
Here, $P(k)\propto E(k)$ is the energy spectrum normalized by the total energy, and $N(k)$ can be viewed as the probability (or number fraction) associated with the scale $k$.

To model the cascade, we interpret $P(k)$ as the distribution of fictitious particles X in k-space, each carrying an infinitesimal energy $dE=u_0^2/2N$ with total number of particles $N\rightarrow \infty$. The nonlinear energy transfer is represented as a one-dimensional continuous-time random walk of these energy carriers: a local jump of X from $k\rightarrow k+\delta k$ corresponds to transferring the infinitesimal energy $dE$ from $k\rightarrow k+\delta k$ due to a local triad interaction at scale $k$ (Fig. \ref{fig:1}). In direct analogy with ordinary diffusion, where spatial random walks generate the evolution of particle density, $P(k,t)$ describes the temporal evolution of the energy distribution in k-space, with a stationary spectrum $P(k)$. This construction naturally incorporates four key properties:
\begin{enumerate}
\item \noindent \textit{Scale-locality.} The energy flux at wavenumber $k$ is assumed to be dominated by local triadic interactions among three modes with $|\textbf{k}|\sim |\textbf{p}|\sim |\textbf{q}|$, while strongly nonlocal interactions are subleading \citep{EYINK:2005-Locality-of-turbulent-cascades,Aluie:2010-Scale-Locality-of-Magnetohydrodynamic-Turbulence}. In the random-walk picture, this is implemented by restricting particles moving only between neighboring shells (k to k+$\delta$k) and energy is transferred only between adjacent scales.
\\
\item \noindent \textit{Energy conservation.} All triadic interactions are energy-conserving. In the absence of forcing and dissipation, this means the total number of particles is conserved. Since each particle carries an infinitesimal energy $dE$, the stochastic dynamics merely redistributes energy in k-space without changing the total energy, implementing conservative transfer across the entire range.
\\
\item \noindent \textit{Viscous dissipation.} Viscosity removes kinetic energy at a rate of $2\nu k^2E(k)$. Translating this into the particle picture produces a scale-dependent sink term $2\nu k^2P(k)$ (Eq.~\eqref{eq:2-5}), which represents the loss of particles (and hence energy) from the shell $k$ due to the viscous damping of modes $k$. The $k^2$ dependence reflects the Laplacian in the Navier-Stokes equations and ensures that dissipation is negligible at large scales and dominant on small scales. \added{We emphasize that this sink term is imported directly from the Navier-Stokes dissipation rate rather than derived from the random-walk construction itself: the random-walk/Fokker-Planck formalism specifies how particles are transported \emph{within} once a waiting time $\tau_{gk}(k)$ is given, but it does not, by itself, constrain the functional form of an external loss term.}
\\
\item \noindent \textit{Scale-dependent waiting time.} The waiting time $\tau_{gk}(k)$ between successive steps for each X particle is identified with the characteristic lifetime $\tau_k(k)$ of eddies at scale $k$. In Kolmogorov phenomenology, this lifetime is a power law of $k$, and, without loss of generality, we parameterize it as $\tau_{gk}(k)\propto k^{-{\gamma}}$ with $\gamma>0$. This \added{is the simplest form that admits a self-similar solution} and captures the fact that smaller-scale eddies (larger $k$) interact and decorrelate more rapidly. It provides a minimal parametrization of possible deviations from the Kolmogorov scaling when intermittency or additional physics are included. The exponent $\gamma$ will be determined self-consistently below.
\end{enumerate}

We consider an idealized initial condition in which all X particles are concentrated at k=0, corresponding to energy injection exclusively at the largest scales ($k_L$=0). In this limit, there is no distinct energy-containing range and only the universal equilibrium range is relevant. The subsequent evolution of the particle distribution in k-space is modeled as a one-dimensional continuous-time random walk with a position-dependent residence (waiting) time $\tau_{gk}(k)$. The corresponding Langevin equation in k-space reads
\begin{equation} 
\label{eq:2-3} 
\frac{d\boldsymbol{k}_{t} }{dt} = \sqrt{2 D_P(\boldsymbol{k}_t)} \xi\left(t\right)\propto \frac{\delta k}{\tau_{gk}}\propto \frac{\delta k}{k^{-\gamma}},    
\end{equation} 
where $\xi(t)$ is Gaussian white noise. Here, $\delta k/\tau_{gk}(k)$ denotes the average rate of change of $k_t$. We assume a power-law scaling $\tau_{gk}(k)\propto k^{-\gamma}$, implying a position-dependent k-space diffusivity
\begin{equation} 
\label{eq:2-4} 
D_{P}(\boldsymbol{k}_t)=D_{0} k^{2\gamma},         
\end{equation} 
where $D_{0}$ is a proportional constant. Thus, smaller $k$ corresponds to reduced k-space diffusivity and consequently a longer waiting time.

The evolution of the particle distribution $P(k,t)$ is governed by the corresponding Fokker-Planck equation
\begin{equation} 
\label{eq:2-5} 
\frac{\partial P(k,t)}{\partial t} = \underbrace{\frac{\partial }{\partial k}\left[ \alpha\frac{\partial D_P}{\partial k}P +D_P \frac{\partial P}{\partial k}\right]}_{1}-\underbrace{2\nu k^2P}_{2},
\end{equation} 
where the first term represents energy transfer in k-space generated by the Langevin Eq.~\eqref{eq:2-3}, with a position-dependent diffusivity $D_P(k)$ and an interpretation parameter $\alpha$ specifying the stochastic calculus ($\alpha$=1 for It\^{o}, $\alpha$=1/2 for Stratonovich \citep{Sokolov:2010-Ito-Stratonovich-Hänggi}, and other values determined by the underlying microphysics). The second term describes viscous dissipation, leading to a $k$-dependent decay of the X population. The derivation of the $k$-space Fokker-Planck equation can be related to an effective Inertial-range \added{Eddy-Damped Quasi-Normal Markovian (EDQNM)} Closure (Appendix \ref{sec:appen2}), similar to the Leith diffusion model \citep{Leith:1967-Diffusion-Approximation}.

The exact steady state solution of Eq.~\eqref{eq:2-5} reads (Appendix \ref{sec:app})
\begin{equation} 
\label{eq:2-6} 
\begin{split}
P(k)&={A}k^{-(1+\gamma)}\exp\left[-\sqrt{\frac{2\nu}{D_0}}\left(\frac{{k^{2-\gamma}}}{2-\gamma}\right)\right] \\
& = {A}k^{-(1+\gamma)}\exp\left[-\lambda\left(k/k_{\eta}\right)^{2-\gamma}\right],\\
\end{split}
\end{equation} 
where $A$ is a normalization constant. The stochastic interpretation parameter $\alpha$ is uniquely fixed by the solution of the stationary distribution (Appendix \ref{sec:app} and \ref{sec:appen2}) to be $\alpha =(1+1/\gamma)/2$; this is not a free choice but a mathematical constraint ensuring that the steady-state solution Eq. \eqref{eq:2-6} exists. Identifying $P(k)$ with the energy spectrum, we obtain an inertial-range scaling $E(k)\propto k^{-(1+\gamma)}$ at small $k$, controlled by the waiting-time exponent $\gamma$, together with an inertial-dissipation crossover described by an exponential cutoff $\exp(-k^{2-\gamma})$. We introduce a characteristic wavenumber $k_{\eta}$ that marks the onset of exponential decay for $k>k_{\eta}$. This allows us to define a dimensionless parameter $\lambda$
\begin{equation}
\label{eq:2-7} 
\lambda=\sqrt{\frac{2\nu}{D_0}}\frac{k_{\eta}^{2-\gamma}}{2-\gamma}.
\end{equation} 
The $n$th moment of the distribution $P(k)$ can be easily found as 
\begin{equation} 
\label{eq:2-8} 
\int_{k_L}^{\infty}P(k)k^ndk=\frac{Ak_{\eta}^{n-\gamma}}{(2-\gamma)\lambda^{\frac{n-\gamma}{2-\gamma}}}\Gamma\left(\frac{n-\gamma}{2-\gamma},\lambda\left(\frac{k_L}{k_{\eta}}\right)^{2-\gamma}\right),
\end{equation} 
where $k_L$ denotes the integral wavenumber scale. Here, $\Gamma(x,y)$ is the upper incomplete Gamma function with limits 
\begin{equation} 
\label{eq:2-8-2} 
\lim_{y \to 0} \Gamma(x,y) = \Gamma(x)-y^x/x  \quad \textrm{and} \quad \lim_{y \to \infty} \Gamma(x,y) = 0.
\end{equation} 

The constant $A$ is obtained by imposing $\int_{k_L\rightarrow 0}^{\infty}P(k)dk=1$ (zeroth moment with n=0 in Eq.~\eqref{eq:2-8}), 
\begin{equation}
\label{eq:2-9} 
A=\frac{(2-\gamma)\lambda^{-\frac{\gamma}{2-\gamma}}k_{\eta}^{\gamma}}{\Gamma\left(-\frac{\gamma}{2-\gamma},\lambda\left(\frac{k_L}{k_{\eta}}\right)^{2-\gamma}\right)}\approx\frac{\gamma k_L^{\gamma}}{1+\frac{\gamma}{2-\gamma}\Gamma\left(-\frac{\gamma}{2-\gamma}\right)\lambda^{\frac{\gamma}{2-\gamma}}\left(\frac{k_L}{k_{\eta}}\right)^{\gamma}}
\end{equation} 
and for $k_L\ll k_{\eta}$, simplifies to $A\approx \gamma k_L^{\gamma}$. In this limit ($k_L\rightarrow 0$), the distribution finally reduces to
\begin{equation}
\label{eq:2-10} 
P(k)=\frac{2E(k)}{u_0^2}=\gamma k_L^{\gamma}k^{-(1+\gamma})\exp\left[-\lambda\left(\frac{k}{k_{\eta}}\right)^{2-\gamma}\right].
\end{equation} 
The integral and dissipation wavenumbers, $k_L$ and $k_{\eta}$, are related to the corresponding real-space integral scale $L$ and dissipative scale $\eta$ by a constant $\beta=1/({2\pi})$,
\begin{equation}
\label{eq:2-11} 
{\eta}={1}/({\beta k_{\eta}}) \quad \textrm{and} \quad L={1}/({\beta k_L}).
\end{equation} 

We next introduce the standard spectral kinetic energy flux function $\Pi_E(k)$ defined as the net transfer of energy from all modes smaller than $k$ to all modes larger than $k$. Using $E(k)$ in Eq.~\eqref{eq:2-10}, the energy flux can be written as
\begin{equation}
\label{eq:2-12} 
\begin{split}
\Pi_E(k)&=\int_k^{\infty}2\nu E(k')k'^2dk'\\
&=\frac{\gamma\nu u_0^2k_{\eta}^2}{(2-\gamma)\lambda}\left(\frac{k_L}{k_{\eta}}\right)^{\gamma} \exp\left[-\lambda\left(\frac{k}{k_{\eta}}\right)^{2-\gamma}\right]
\end{split}
\end{equation} 
and is related to the spectrum $E(k)$ through
\begin{equation}
\label{eq:2-13} 
\Pi_E(k)=\frac{2\nu}{(2-\gamma)\lambda} k_{\eta}^{2-\gamma}k^{1+\gamma}E(k).
\end{equation} 

It is important to note that in the inertial range $k_L\ll k\ll k_{\eta}$, the exponential factor in Eq. \eqref{eq:2-12} approaches unity, so that $\Pi_E(k)\approx \mathrm{const}$ exactly. The $k$-independence of the energy flux is therefore not assumed but is a rigorous consequence of the steady-state Fokker-Planck equation in that range. The mean dissipation rate $\varepsilon$ follows from the limit $k\rightarrow 0$,
\begin{equation}
\label{eq:2-14} 
\begin{split}
\varepsilon=\Pi_E(k\rightarrow0)&=\frac{\gamma\nu u_0^2}{(2-\gamma)\lambda} k_L^{\gamma}k_{\eta}^{2-\gamma}\\
&=\frac{\gamma}{2-\gamma}\frac{1}{\lambda\beta^2}\left(\frac{u_0^{2/\gamma}}{L}\right)^{\gamma}\left(\frac{u_{\eta}^{1/(1-\gamma)}}{\eta}\right)^{1-\gamma},
\end{split}
\end{equation} 
where $u_{\eta}={\nu}/{\eta}$ is the characteristic velocity at scale $\eta$. The magnitude of $\varepsilon$ is set by the boundary conditions ($u_0$ and $L$ and $u_{\eta}$ and $\eta$); the framework predicts the functional relationship between $\varepsilon$ and the spectral shape, rather than $\varepsilon$ in isolation.

We have so far characterized the turbulence spectrum in terms of a scale-dependent waiting time $\tau_{gk}\propto k^{-\gamma}$ with an a priori unknown exponent $\gamma$. \added{We stress that the stochastic transport formalism developed above (Eqs.~\eqref{eq:2-3}-\eqref{eq:2-10}) does not by itself fix $\gamma$: any power-law waiting time yields a self-consistent stationary Fokker-Planck solution, Eq.~\eqref{eq:2-6}, for arbitrary $\gamma$. Determining the physical value of $\gamma$ therefore requires one additional, physically motivated closure relation that connects the waiting time $\tau_{gk}(k)$ to the local eddy dynamics; this closure is external to the stochastic-transport itself, in the same way that the value of $\lambda$ in the halo random walk  (Section~\ref{sec:3}) requires input from either the primordial spectrum or gravitational dynamics.} 

To fix $\gamma$, we invoke a simple physical argument. For a spectrum $E(k) \propto k^{-(1+\gamma)}$, the energy associated with wavenumber k scales as $u_k^2\propto E(k)k\propto k^{-\gamma}$, so that the characteristic eddy velocity satisfies $u_k \propto k^{-\gamma/2}$. The corresponding eddy-turnover time is then $\tau_k \propto (u_kk)^{-1}\propto k^{\gamma/2-1}$\added{; this local-interaction-time estimate is used here only to fix $\gamma$}. In our random-walk picture, \added{an additional physically motivated closure assumption is:} the transfer of an infinitesimal energy element $dE$ from k to $k+\delta k$ is mediated by local triadic interactions, whose characteristic time (waiting time $\tau_{gk}$) is proportional to eddy turn-over time $\tau_k$. Thus the waiting time must satisfy $\tau_{gk}\propto \tau_k\propto k^{\gamma/2-1}$. Consistency with scaling $\tau_{gk} \propto k^{-\gamma}$ requires
\begin{equation}
\label{eq:2-15} 
\frac{\gamma}{2}-1=-\gamma \quad \textrm{such that}\quad \gamma=2/3,
\end{equation} 
where $\gamma=2/3$ is the unique value for which the Langevin equation \eqref{eq:2-3} is self-consistent with the nonlinear transfer rate derived from the spectrum it generates. \added{We emphasize that this is a self-consistency condition built on the closure $\tau_{gk}\propto\tau_k$ introduced above, rather than a result derived from the stochastic transport equations alone.} 

We note that the self-consistent exponent $\gamma=2/3$ corresponds to a velocity field of H\"older regularity $h=1/3$ ($E(k)\propto k^{-(1+2h)}$), which is precisely the critical threshold identified by Onsager's conjecture \citep{Onsager:1949-Statistical-hydrodynamics}: weak Euler solutions with $h>1/3$ conserve energy, while $h\le1/3$ permits anomalous dissipation in the inviscid limit, as established rigorously via convex integration \citep{Isett:2018-A-proof-of-Onsagers-conjecture}. This is an independent, purely kinematic/PDE-regularity result, distinct from the present stochastic-transport closure; both frameworks single out the same critical exponent, which lends further support to $\gamma=2/3$ as a physically robust fixed point.

While the final result reproduces the K41 scaling, we do not invoke a scale-independent $\varepsilon$ as a priori hypothesis; instead, $\varepsilon$ emerges from the steady-state dynamics (Eq. \eqref{eq:2-14}) \added{once the closure fixing $\gamma$ is supplied}. With $\gamma=2/3$, Eq.~\eqref{eq:2-10} reproduces the Kolmogorov -5/3 spectrum together with an exponential cutoff $\exp(-k^{4/3})$ across the inertial-dissipation transition. The dissipation rate in Eq.~\eqref{eq:2-14} becomes
\begin{equation}
\label{eq:2-16} 
\varepsilon=\frac{1}{2\lambda\beta^2}\left(\frac{u_0^3}{L}\right)^{2/3}\left(\frac{u_{\eta}^3}{\eta}\right)^{1/3}.
\end{equation} 
Since the rate of energy dissipation is finite, this requires $L\propto u_0^3$ with the limit $k_L\rightarrow 0$ or $L\rightarrow \infty$. For $\gamma=2/3$, the velocity $u_k\propto k^{-\gamma/2}\propto k^{-1/3}$ and $u_k^3k$ is a constant. This is equivalent to a relation $u_0^3/L\sim u_k^3/l\sim u_{\eta}^3/\eta$ for scales $\eta<l<L$, where $u_k$ is the characteristic velocity at wavenumber $k$ or scale $l$ in real space. We now choose a constant $C_X$ and the scale $\eta$ to satisfy an exact relation
\begin{equation}
\label{eq:2-17} 
\varepsilon=C_X\frac{u_0^3}{L}=\frac{u_{\eta}^3}{\eta} \quad \textrm{and} \quad C_X^{2/3} \cdot 2\lambda\beta^2=1,
\end{equation} 
where $C_X$ is a dimensionless cascade efficiency: the fraction of large-scale kinetic energy that cascades to dissipation per large-eddy turnover time. Direct numerical simulations (DNS) of turbulence at high Reynolds number consistently yield $C_X\approx 0.4$--$0.6$, independently of the forcing details \citep{Kaneda:2003-Energy-dissipation-rate}. 

The value of $C_X$ is not a free fitting parameter but can be independently constrained from Kolmogorov's exact 4/5 law \citep{Kolmogoroff:1941-Dissipation-of-energy-in-the-l}, which is derivable directly from the Navier--Stokes equations without closure assumptions: $\langle(\delta u_L)^3\rangle = -\tfrac{4}{5}\varepsilon r$. Using $\varepsilon = C_X u_0^3/L$ gives $\langle(\delta u_L(L))^3\rangle/u_0^3 = -4C_X/5$. The left-hand side is the velocity-increment skewness at the integral scale, measured experimentally as $|S_3(L)|\approx 0.4$-$0.5$ \citep{Sukoriansky:2018-Inertial-range-skewness}, which yields $C_X = 5|S_3|/4\approx 0.5$--$0.6$, entirely consistent with DNS. 
With Eq. \eqref{eq:2-17} and $u_{\eta}=\nu/\eta$, the Kolmogorov microscales naturally emerge as $\eta=(\nu^3/\varepsilon)^{1/4}$.

Using Eqs. \eqref{eq:2-10} and \eqref{eq:2-17}, the full equilibrium-range spectrum and the Kolmogorov constant $C_K$ are
\begin{equation}
\label{eq:2-18} 
\begin{split}
&E(k)=C_K \varepsilon^{2/3}k^{-5/3}\exp\left[-\frac{3}{2}C_K(\eta k)^{4/3}\right],\\
& C_K=\frac{1}{3}\left(\beta C_X\right)^{-2/3} \quad \textrm{and} \quad \lambda=\frac{3}{2}C_K\beta^{-4/3}.
\end{split}
\end{equation} 
Thus, the empirical Pao spectrum in Eq. \eqref{eq:1-1} is analytically recovered without invoking any additional assumptions in the original derivation \cite{Pao:1965-Structure-of-Turbulent-Velocity}. With $C_K\approx 1.6$ for $C_X\approx 0.5$-$0.6$ (consistent with DNS measurements), the theoretical contribution of Eq. \eqref{eq:2-18} is the functional relation $C_K = (\beta C_X)^{-2/3}$/3 (with $\beta=1/(2\pi)$), which constrains the universal spectrum constant $C_K$ in terms of the measurable cascade efficiency $C_X$, without any ad hoc fitting. The predicted spectrum Eq. \eqref{eq:2-18} with $C_K\approx 1.6$ (for $C_X\approx 0.5$--$0.6$) is in quantitative agreement with published measurements. DNS at Taylor-scale Reynolds numbers $Re_\lambda \approx 1000$--$2000$ \citep{Kaneda:2003-Energy-dissipation-rate, Ishihara:2009-Small-scale-statistics} consistently yield $C_K = 1.5$--$1.7$. The exponential dissipation-range cutoff $\exp[-\frac{3}{2}C_K(\eta k)^{4/3}]$ is also consistent with measured spectra across a wide range of Reynolds numbers, confirming the functional form of Eq.~\eqref{eq:2-18} beyond the inertial range. Grid-turbulence experiments \citep{Sreenivasan:1995-On-the-universality-of-the-Kolmogorov} report $C_K = 1.5\pm 0.1$, in good agreement.

In this framework, a random walk of energy-carriers in wavenumber space, driven by local triadic interactions, suffices to reproduce the complete equilibrium energy spectrum, suggesting a promising avenue for extensions to MHD turbulence and intermittency. Appendix \ref{sec:appen2} provides a derivation of the inertial-range random walk and Fokker-Planck model from an effective EDQNM closure. Appendix \ref{sec:comparison} and Table \ref{table:C1} provide a detailed comparison between the current stochastic random-walk framework and Kolmogorov’s K41 theory. In the following sections, the same approach is extended to cosmic structure formation.

\section{Halo random walk for mass function}
\label{sec:3}

\begin{figure}
\includegraphics*[width=\columnwidth]{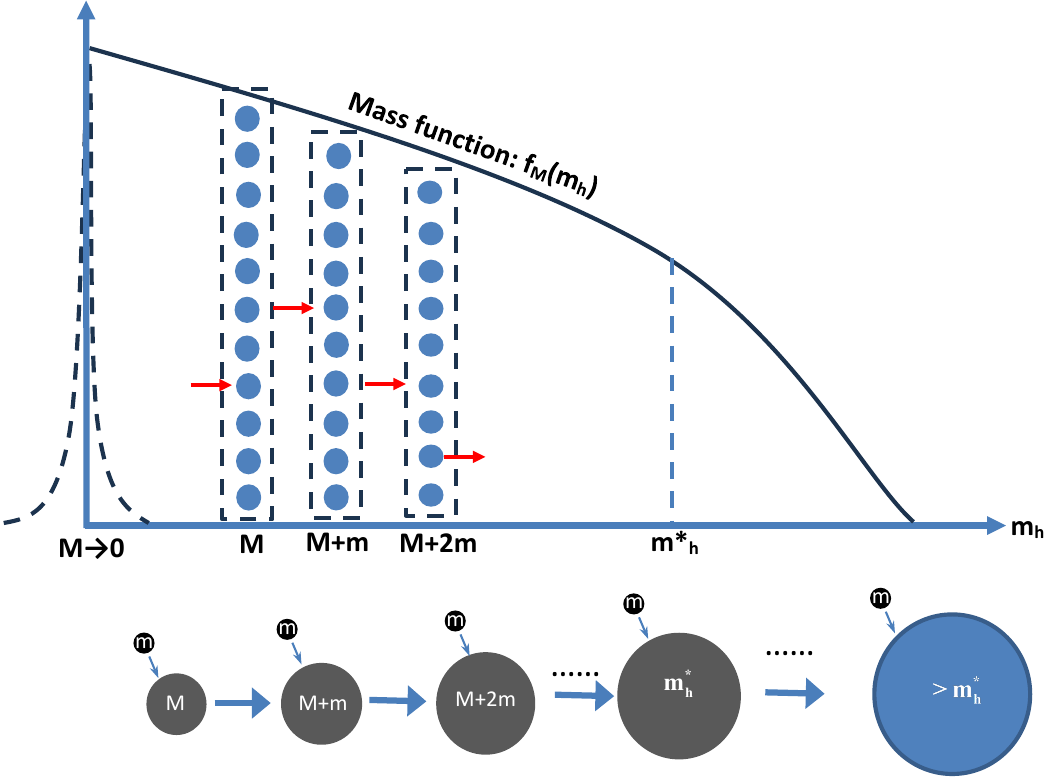}
\caption{Schematic of the halo random walk in mass space. A halo of mass $M$ that accretes a free particle of mass $m$ steps to $M+m$, carrying mass toward larger halo masses. For a halo of mass $m_h$, such events occur after an average waiting time $\tau_g(m_h)$. Starting from sharply localized initial conditions, the process generates the halo mass function as the solution of the associated Fokker--Planck equation.}
\label{fig:S2}
\end{figure}

For collisionless dark matter, long-range gravity favors a broad halo mass spectrum that maximizes the entropy of the self-gravitating system \citep{Xu:2023-Maximum-entropy-distributions-of-dark-matter}. Haloes are the basic elements of the flow, analogous to eddies in turbulence. They mediate the transfer of mass and energy across scales, and their abundance is encoded in the halo mass function. Hierarchical structure formation can therefore be viewed as a random walk in mass space (Fig.~\ref{fig:S2}). A halo of mass $M$ that accretes a particle of mass $m$ steps to $M+m$, and repeated steps generate a continuous flux of mass toward larger masses. For haloes of mass $m_h$, these events occur after an average waiting time $\tau_g(m_h)$. Following the Lagrangian equation~\eqref{eq:2-3} for the turbulence spectrum, we take $\tau_g\propto m_h^{-\lambda}$, with $\lambda$ set by the initial spectrum or by gravity, whichever dominates; (this exponent is unrelated to the dimensionless constant $\lambda$ of Eq.~\eqref{eq:2-7}; similarly, $\alpha$ and $\beta$ denote different quantities in different sections). The population starts sharply localized at $m_h=0$, representing many small, dense proto-haloes at high redshift.

\paragraph*{Fokker--Planck description.}
The halo mass performs a random walk along a clock $C(a)$, specified below. It obeys the Langevin equation \citep{Xu:2023-Dark-matter-halo-mass-functions-and}
\begin{equation}
\label{eq:3-1}
\frac{\partial m_h}{\partial C}=\sqrt{2D_p(m_h)}\,\varsigma(C),\qquad
\langle\varsigma(C)\varsigma(C')\rangle=\delta(C-C').
\end{equation}
Massive haloes grow faster because their potential wells are deeper. The waiting time $\tau_g\propto m_h^{-\lambda}$ therefore implies an rms mass increment per unit clock $\propto m_h^{\lambda}$, i.e.\ the diffusivity
\begin{equation}
\label{eq:3-2}
D_p(m_h)=D_{p0}\,m_h^{2\lambda}.
\end{equation}
The mass function $f_M(m_h,a)$, the distribution of halo mass obeys
\begin{equation}
\label{eq:3-3}
\frac{\partial f_M}{\partial C}=\frac{\partial}{\partial m_h}\left[\alpha\frac{\partial D_p}{\partial m_h}f_M+D_p\frac{\partial f_M}{\partial m_h}\right],
\end{equation}
where $\alpha$ specifies the interpretation of the stochastic integral. For the Stratonovich choice $\alpha=1/2$ (Appendix~\ref{sec:app}), Eq.~\eqref{eq:3-3} has the closed-form solution
\begin{equation}
\label{eq:3-4}
f_M(m_h,a)=\frac{m_h^{-\lambda}}{\sqrt{\pi D_{p0}C}}\exp\left[-\frac{m_h^{2-2\lambda}}{4(1-\lambda)^2D_{p0}C}\right].
\end{equation}
This is a stretched Gaussian, with a power law at low mass and an exponential cutoff at high mass \citep{Stratonovich:1966-A-new-representation-for-stoch,Metzler:2014-Anomalous-diffusion-models-and}. The Stratonovich convention is physically motivated: by the Wong--Zakai theorem, the limit of a smooth process with short but nonzero correlation time converges to it. Define the characteristic mass
\begin{equation}
\label{eq:3-14}
M_h^*=\left[\frac{(1-\lambda)^2D_{p0}C}{\eta_0}\right]^{1/(2-2\lambda)},
\end{equation}
with $\eta_0$ a dimensionless constant; $M_h^*$ is identified with $m_h^*$ below and should not be confused with the total halo mass $M_h$. The solution then reads
\begin{equation}
\label{eq:3-15}
f_M(m_h,a)=\frac{1-\lambda}{\sqrt{\pi\eta_0}\,M_h^*}\left(\frac{m_h}{M_h^*}\right)^{-\lambda}\exp\left[-\frac{1}{4\eta_0}\left(\frac{m_h}{M_h^*}\right)^{2-2\lambda}\right].
\end{equation}
The exponent $\lambda$ sets the waiting time, the low-mass slope and the cutoff shape, and all time dependence is carried by $M_h^*$. Equation~\eqref{eq:3-14} is a diffusion front in mass space, $M_h^{*\,2-2\lambda}\propto C$.

\paragraph*{The clock.}
With $y\equiv m_h^{1-\lambda}/(1-\lambda)$, Eqs.~\eqref{eq:3-1}--\eqref{eq:3-2} reduce to free diffusion in $y$, which spreads as $y^2\propto D_{p0}C$. A halo near the front grows with it at the mean rate $dm_h/dt\propto m/\tau_g$, so that,
\begin{equation}
\label{eq:3-clock}
\frac{d\sqrt{C}}{dt}\propto\frac{1}{m_h^{\lambda}\,\tau_g}.
\end{equation}
The right side is mass independent, consistent with $\tau_g\propto m_h^{-\lambda}$. For $\tau_g\propto m_h^{-\lambda}a^{x_\tau}$, Eq.~\eqref{eq:3-clock} gives $C\propto a^{3-2x_\tau}$ in matter domination. Two clocks occur:
\begin{itemize}
\item \emph{Linear clock}, $x_\tau=1/2$. At fixed mass, accretion events are evenly spaced in the growth factor $D=a$, as in linear collapse paced by $\delta\propto D$. Then $C=D^2=a^2$.
\item \emph{Cascade clock}, $x_\tau=1$. Events are evenly spaced in conformal time $\eta\propto a^{1/2}$, as for encounters at a fixed comoving rate and $C=a$.
\end{itemize}
In both cases the clock is the square of the variable in which events are evenly spaced, because $y$ advances linearly in that variable while the front advances as $\sqrt{C}$. 

\paragraph*{Press--Schechter limit.}
For a scale-free linear spectrum $P(k)\propto k^n$, $\sigma_\delta(m_h)\propto m_h^{-(3+n)/6}$, and the peak height is
\begin{equation}
\label{eq:3-15-2}
\nu\equiv\frac{\delta_c^2}{\sigma_\delta^2(m_h,a)}=\left(\frac{m_h}{M_{PS}^*}\right)^{2-2\lambda},\qquad \lambda=\frac{3-n}{6},
\end{equation}
where $M_{PS}^*$, defined by $\sigma_\delta(M_{PS}^*,a)=\delta_c$, plays the role of $M_h^*$. Since $\nu\propto m_h^{(3+n)/3}/D^2$, this limit runs on the linear clock and gives $M_{PS}^*\propto D^{6/(3+n)}$. In terms of $\nu$, Eq.~\eqref{eq:3-15} becomes
\begin{equation}
\label{eq:3-15-3}
f_{PS}(\nu)=\frac{1}{2\sqrt{\pi\eta_0}}\,\nu^{-1/2}\exp\left(-\frac{\nu}{4\eta_0}\right),
\end{equation}
the Press--Schechter form for $\eta_0=1/2$ \citep{Press:1974-Formation-of-Galaxies-and-Clus}.

This reflects a duality with the extended Press--Schechter (EPS) excursion set \citep{Bond:1991-Excursion-Set-Mass-Functions-f}. In EPS, the linear overdensity $\delta$ walks in the variance $S=\sigma_\delta^2(M)$ toward a barrier $\omega=\delta_c/D$, with $\nu=\omega^2/S$. Here $m_h$ walks along $C$, with $\nu\propto m_h^{2-2\lambda}/C$. By the reflection principle, the fraction of mass in halos above mass $M$ equals the probability that an unconstrained walk at fixed $S(M)$ lies beyond the barrier, $F(>M)=P(|\delta|>\omega)=\operatorname{erfc}\sqrt{\nu/2}$. The PS distribution of $\nu=\omega^2/S$ is therefore the distribution of $\delta^2/S$ for a free Gaussian folded at zero. Equation~\eqref{eq:3-4} has the same form in $y$, folded by the reflecting boundary at $m_h=0$, which supplies the PS factor of two. Matching mass with mass and time with time gives $m_h^{2-2\lambda}\leftrightarrow S^{-1}$, i.e.\ $\lambda=(3-n)/6$, and $C\leftrightarrow\omega^{-2}\propto D^2$.

EPS therefore encodes only initial-condition statistics: its time enters through the barrier alone, and its clock is always $D^2$. The present model can also run on the cascade clock and so describe nonlinear redistribution. For a CDM spectrum, $n$ varies with scale and $\lambda$ is a local effective parameter.

\begin{figure}
\includegraphics[width=\columnwidth]{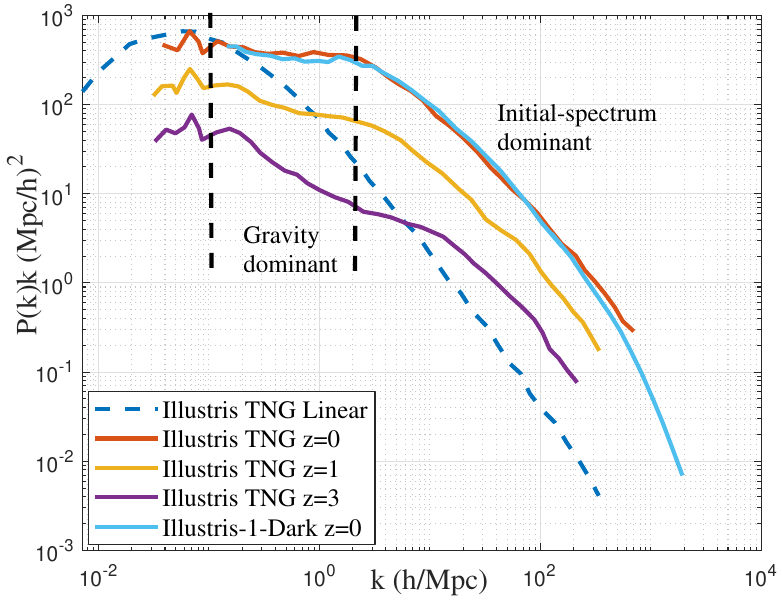}
\caption{Compensated density power spectrum $P(k)k$ for IllustrisTNG \citep{Springel:2018-First-results-from-the-IllustrisTNG} and Illustris-1-Dark \citep{Vogelsberger:2014-Introducing-the-Illustris-Project} at $z=0,\,1,\,3$, with the linear prediction. A horizontal plateau indicates $n=-1$, i.e.\ $P(k)\propto k^{-1}$, as expected for gravity-dominated dynamics. The plateau on 1--10\,Mpc scales (dashed lines) widens toward lower redshift as nonlinear collapse advances; on sub-Mpc scales the spectrum declines, retaining the primordial imprint.}
\label{fig:3-1}
\end{figure}

\paragraph*{Characteristic mass $m_h^*$.}
Near the characteristic mass $m_h^*$, repeated mergers and smooth accretion erase the memory of the initial conditions, and gravity sets the exponent. The compensated spectrum $P(k)k$ shows this directly (Fig.~\ref{fig:3-1}). A universal plateau with $n=-1$ develops around 1\,Mpc and extends to larger scales with time, while sub-Mpc scales retain the primordial imprint \citep{Xu:2021-Inverse-mass-cascade-mass-function}.

\begin{figure}
\includegraphics[width=\columnwidth]{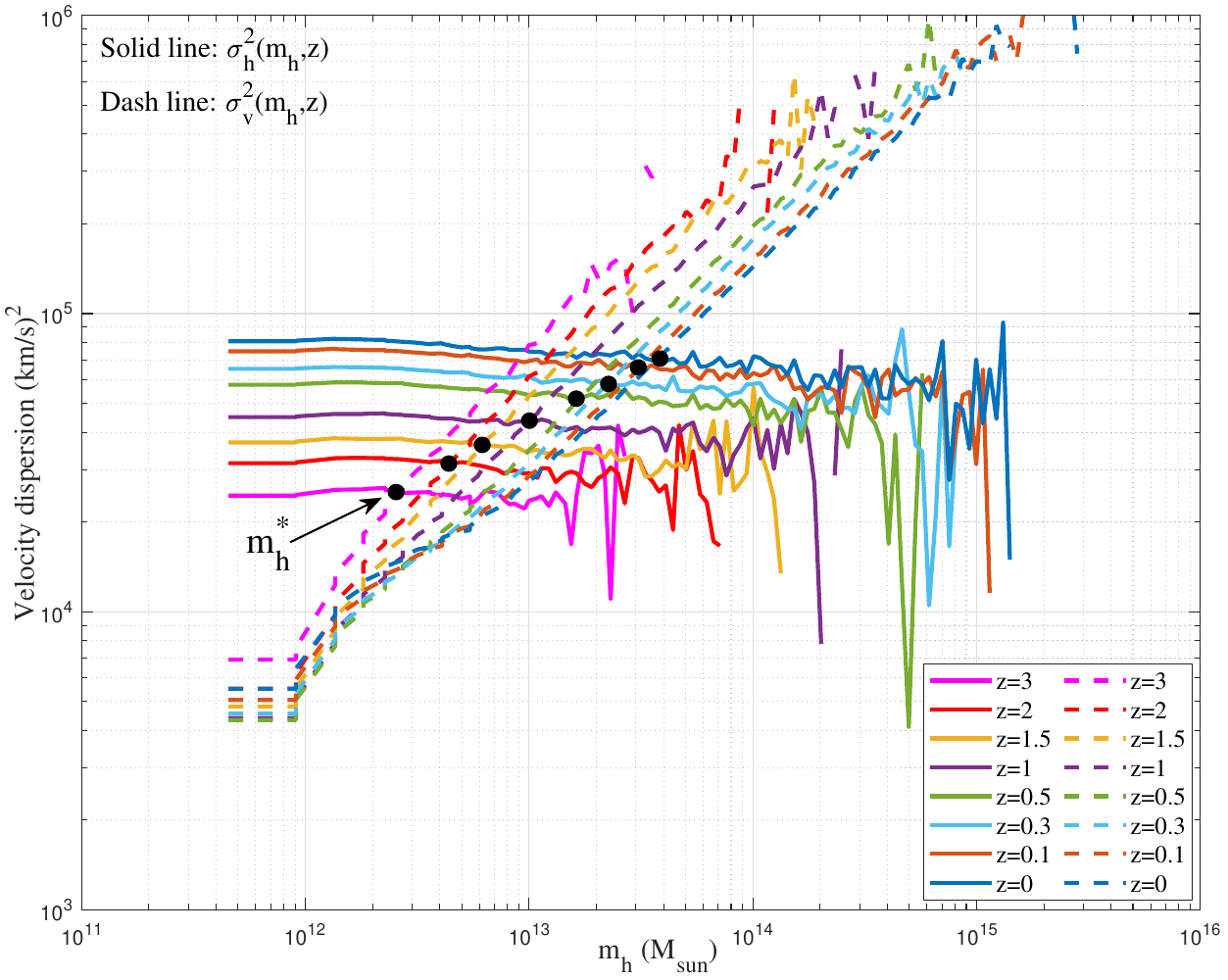}
\caption{Virial dispersion $\sigma_v^2(m_h,a)$ (one-dimensional particle dispersion within haloes of mass $m_h$) and bulk dispersion $\sigma_h^2(m_h,a)$ (one-dimensional dispersion of halo centre-of-mass velocities) versus halo mass at several redshifts in the Virgo SCDM simulation \citep{Jenkins:1998-Evolution-of-structure-in-cold}. The former rises steeply with mass; the latter is nearly mass independent. Their intersections (black dots) define $m_h^*(a)$, whose evolution is shown in Fig.~\ref{fig:4}.}
\label{fig:3}
\end{figure}

We define $m_h^*$ by the kinematic crossover (Appendix~\ref{sec:chara_mass})
\begin{equation}
\label{eq:velocity_matching}
\sigma_v^2(m_h^*,a)=\sigma_h^2(m_h^*,a),
\end{equation}
where $\sigma_v^2$ is the one-component virial dispersion inside haloes and $\sigma_h^2$ the dispersion of halo bulk velocities (Fig.~\ref{fig:3}). The bulk flow is driven by super-halo perturbations,
\begin{equation}
\label{eq:bulk_velocity_scaling}
\sigma_h^2\propto\int_0^\infty P(k,a)\,W^2(kR)\,dk\propto a^2H^2R^2\sigma_\delta^2(m_h,a),
\end{equation}
for $m_h\gtrsim m_h^*$, and is nearly independent of $m_h$ below $m_h^*$, where it saturates (Appendix~\ref{sec:chara_mass}). At a given time, haloes near $m_h^*$ share one dynamical density, so
\begin{equation}
\label{eq:virial_scaling}
\sigma_v^2(m_h,a)\propto\frac{Gm_h}{r_{\rm vir}}\propto m_h^{2/3},
\end{equation}
with $aR/r_{\rm vir}\simeq\Delta_{\rm vir}^{1/3}$; this density tracks a fixed overdensity $\Delta_{\rm vir}\simeq18\pi^2$ only before $a_*$. The dynamical peak height is therefore
\begin{equation}
\label{eq:nu_dyn}
\nu\simeq\frac{\sigma_v^2(m_h,a)}{\sigma_h^2(a)}\simeq\left(\frac{m_h}{m_h^*}\right)^{2/3},
\end{equation}
which by Eq.~\eqref{eq:3-15-2} corresponds to
\begin{equation}
\label{eq:nu_dyn-2}
\lambda=\frac{2}{3},\qquad n=-1,
\end{equation}
the slope of the plateau in Fig.~\ref{fig:3-1}. Because it follows from the virial law, this exponent holds for $m_h^*$ in every regime.

\begin{figure}
\includegraphics[width=\columnwidth]{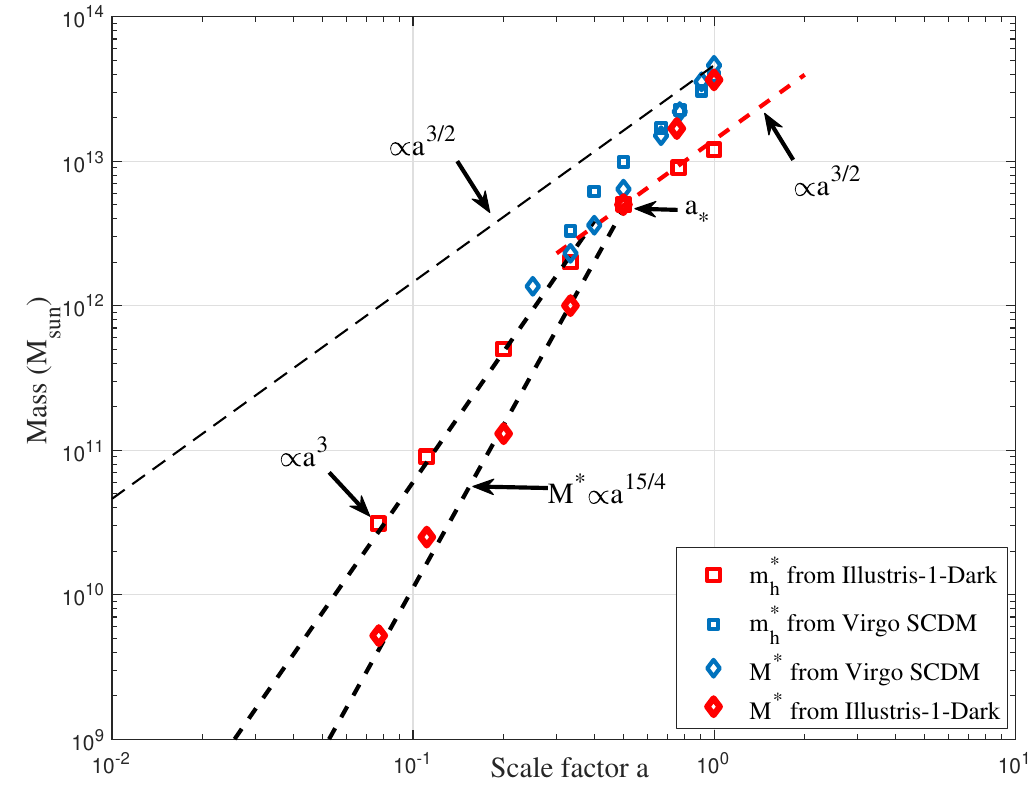}
\caption{Evolution of two mass scales, $M^*(a)$ and $m_h^*(a)$, in the Virgo SCDM \citep{Jenkins:1998-Evolution-of-structure-in-cold} and Illustris-1-Dark \citep{Vogelsberger:2014-Introducing-the-Illustris-Project} simulations. In Illustris-1-Dark, $m_h^*\propto a^{3}$ ($M^*\propto a^{15/4}$) before the crossover $a_*$ and $m_h^*\propto a^{3/2}$ ($M^*\propto a^3$) at late times (dashed line).}
\label{fig:4}
\end{figure}

The mass $m_h^*$ is distinct from $M_{PS}^*$. The latter is a linear-theory estimate, the mass at which the linearly extrapolated field reaches $\delta_c$, whereas $m_h^*$ is the dynamical scale at which bulk and virial kinetic energies are equal. In Illustris-1-Dark, $m_h^*\propto a^3$ for $a\lesssim a_*$ and approaches $m_h^*\propto a^{3/2}$ for $a\gtrsim a_*$, with a crossover time $a_*\simeq 0.5$ (Fig.~\ref{fig:4}). The PS mass grows faster, so $m_h^*/M_{PS}^*$ falls from $\sim50$ at $z\approx2$ to order unity today.

\paragraph*{Waiting time near $m_h^*$.}
On the cascade clock $C=a$, Eq.~\eqref{eq:3-14} with $\lambda=2/3$ gives $m_h^*\propto a^{3/2}\propto t$: the characteristic mass grows at a constant rate. The waiting time is then
\begin{equation}
\label{eq:tau_g}
\tau_g(m_h,a)\propto m_h^{-2/3}\,a,
\end{equation}
corresponding to $x_\tau=1$ in Eq.~\eqref{eq:3-clock}, so that $\tau_g(m_h^*)$ is constant. Equation~\eqref{eq:tau_g} applies near $m_h^*$. The mass dependence has a direct interpretation. Haloes near $m_h^*$ share one dynamical density, so $r_h\propto m_h^{1/3}$, and a waiting time inversely proportional to the potential gives
\begin{equation}
\label{eq:tau_phi}
\tau_g\propto\frac{1}{\Phi}\propto\frac{r_h}{Gm_h}\propto m_h^{-2/3}:
\end{equation}
haloes with deeper potentials accrete faster. The factor $a$ is not set by $\Phi$, since haloes near $m_h^*$ do not keep a fixed contrast in this regime (Table~\ref{tab:two_scales}). Instead it states that the waiting time is constant in conformal time: encounters are diluted by expansion.

\paragraph*{Double-$\lambda$ mass function.}
A single $\lambda$ cannot describe the full mass range, because the slope and the cutoff of the mass function are set by different physics. Below the cutoff, mass flows through a power-law range of slope $\lambda_1$ that still encodes the memory of primordial spectrum. The cutoff itself is shaped by a local gravity-dominant $\lambda_2$ there. We define the cutoff $M^*$ by matching virial kinetic energies,
\begin{equation}
\label{eq:ts_Mstar}
\sigma_v^2(M^*,a)=\langle u^2\rangle/W,
\end{equation}
where $\langle u^2\rangle=\tfrac23\langle K_{pv}\rangle$ is the mass-weighted virial dispersion of all halo mass and $W=O(1)$ is a shape factor. Because the halo mass per logarithmic interval rises toward the cutoff ($\propto m_h^{1-\lambda_1}$ with $\lambda_1<1$), haloes near $M^*$ set the mean virial energy. Like $m_h^*$, $M^*$ is a nonlinear quantity, distinct from $M_{PS}^*$.

The single-$\lambda$ form~\eqref{eq:3-15} generalizes to \citep{Xu:2023-Dark-matter-halo-mass-functions-and,Xu:2021-Inverse-mass-cascade-mass-function}
\begin{equation}
\label{eq:3-18}
f_M(m_h,a)=\frac{2(1-\lambda_2)}{\Gamma(q/2)}\frac{(2\sqrt{\eta_0})^{-q}}{M^*}\left(\frac{M^*}{m_h}\right)^{\lambda_1}\exp\left[-\frac{1}{4\eta_0}\left(\frac{m_h}{M^*}\right)^{2-2\lambda_2}\right],
\end{equation}
with $q=(1-\lambda_1)/(1-\lambda_2)$. Here $\lambda_2$ is the cutoff exponent. It equals $2/3$ once $M^*$ lies on the $n=-1$ plateau, i.e.\ after the crossover $a_*$ defined below, and the primordial value $\lambda_i=(3-n_i)/6$ before. In terms of the linear peak height $\nu=(m_h/M_{PS}^*)^{2-2\lambda_i}$, Eq.~\eqref{eq:3-18} takes the three-parameter form
\begin{equation}
\label{eq:3-20}
\begin{split}
&f_{D\lambda}(\nu)=\frac{p\,(2\sqrt{\eta_0^*})^{-q}}{\Gamma(q/2)}\,\nu^{\frac{pq}{2}-1}\exp\left(-\frac{\nu^p}{4\eta_0^*}\right),\\
&p=\frac{1-\lambda_2}{1-\lambda_i},\quad q=\frac{1-\lambda_1}{1-\lambda_2},\quad \eta_0^*=\eta_0\left(\frac{M^*}{M_{PS}^*}\right)^{2-2\lambda_2},
\end{split}
\end{equation}
where $p$ sets the cutoff shape relative to the linear baseline and $q$ compares the slope with the cutoff exponent.

\begin{figure}
\includegraphics[width=\columnwidth]{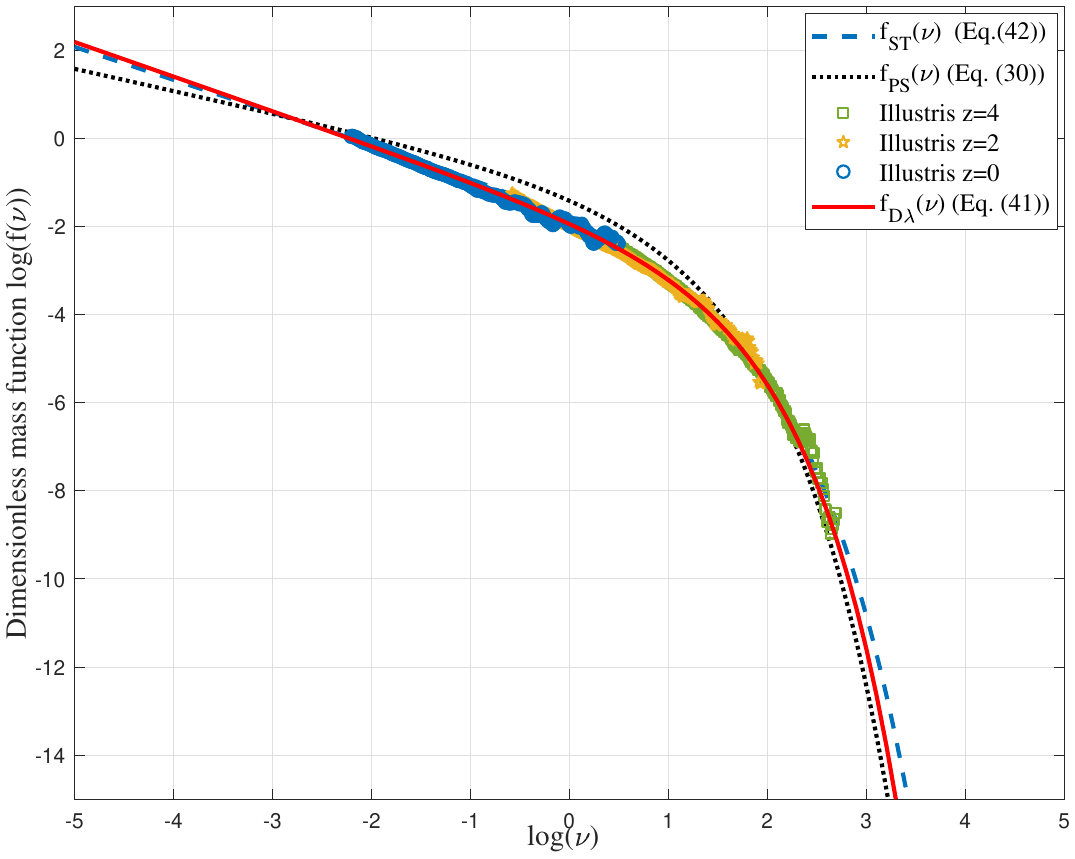}
\caption{Dimensionless halo mass functions $\log f(\nu)$ from Illustris-1-Dark at several redshifts \citep{Vogelsberger:2014-Introducing-the-Illustris-Project}, compared with the PS, ST and double-$\lambda$ (Eq.~\eqref{eq:3-20}) forms. PS overpredicts the abundance near $\nu\approx1$ and underpredicts the slope at $\nu\ll1$. ST fits well overall but overpredicts massive haloes. The double-$\lambda$ function reproduces the simulation across all masses and redshifts.}
\label{fig:4-2}
\end{figure}

Figure~\ref{fig:4-2} compares Eq.~\eqref{eq:3-20} with alternatives. The PS form~\eqref{eq:3-15-3} overestimates the abundance near $\nu\approx1$, because it lacks a gravity-dominated regime. It also underpredicts the slope at $\nu\ll1$, because it neglects the gravitational renormalization of the initial spectral index. The Sheth--Tormen form, which includes ellipsoidal collapse,
\begin{equation}
\label{eq:fST}
f_{ST}(\nu) = \frac{\sqrt{\pi}}{\Gamma(\frac{1}{2})+2^{-p}\Gamma(\frac{1}{2}-p)}\sqrt{\frac{2q}{\pi}} \!\left(1+\frac{1}{(q\nu)^{p}}\right) \frac{e^{-\frac{q\nu}{2}}}{2\sqrt{\nu}},
\end{equation}
with $q=0.75$ and $p=0.3$ \citep{Sheth:2002-An-excursion-set-model-of-hier}, agrees reasonably well but increasingly overestimates massive haloes at high redshift \citep{Reed:2003-Evolution-of-the-mass-function,Lukic:2007-The-Halo-Mass-Function}.

The double-$\lambda$ function fits best, with $\eta_0^*=1.162$, $q=0.365$ and $p=1.185$, i.e.\ $\lambda_1=0.878$, $\lambda_2=2/3$ and $\lambda_i=0.719$. The fit holds these exponents and $\eta_0^*$ fixed at all redshifts, which is an approximation across $a_*$, since $\eta_0^*$ evolves once $M^*$ departs from $M_{PS}^*$. The improvement over $f_{ST}$ at the massive, high-redshift end comes from letting the gravitational cutoff exponent $\lambda_2$ differ from the primordial one $\lambda_i$ ($p\neq1$). Likewise, $q\neq1$ allows the low-mass slope $\lambda_1$, which retains the primordial imprint, to differ from $\lambda_2$, which is set by gravity.

\paragraph*{Two fronts and the crossover $a_*$.}
Both characteristic masses are diffusion fronts of the form of Eq.~\eqref{eq:3-14}, $X^{2-2\lambda}\propto C$. Such a law fixes only the growth $X(a)$, not how it splits into an exponent $\lambda$ and a clock $C$. For each mass, physics fixes one of the two, and the other changes at the crossover $a_*$. For $m_h^*$ the exponent is fixed by the virial law, $\lambda=2/3$, and the clock changes: the linear clock $C=a^2$ before $a_*$ gives $m_h^*\propto a^3$, and the cascade clock $C=a$ after $a_*$ gives $m_h^*\propto a^{3/2}$. 

For $M^*$ the clock is fixed to linear clock, $C=D^2$, because halos beyond $M^*$ are rare peaks of a field growing as $D$, and the exponent changes. Before $a_*$ it follows from the energy cascade. Halos near the cutoff keep a fixed contrast, $\sigma_v^2(M^*)\propto M^{*2/3}a^{-1}$, and $\langle u^2\rangle\propto t$ requires $\sigma_v^2(M^*)\propto a^{3/2}$, so $M^*\propto a^{15/4}$. This gives $\lambda_i=11/15$, i.e.\ a local spectral index $n=-7/5$ at $M^*$, close to the fitted $0.719$. After $a_*$ the cutoff lies on the gravity-dominated plateau, $\lambda_2=2/3$ and $n=-1$, so $M^*\propto a^3$. Energy closure, with halos near $M^*$ no longer at fixed contrast ($\sigma_v^2\propto m_h^{2/3}a^{-1/2}$), independently gives the same result. Hence $M^*/m_h^*\propto a^{3/4}$ rises to unity at $a_*$, and $M^*/m_h^*\propto a^{3/2}$ afterwards. Table \ref{tab:two_scales} summarizes the two regimes.

\textit{The crossover $a_*$} is defined by the balance of two specific kinetic energies, both averaged over halo particles. The bulk energy $\langle K_{ph}\rangle=\tfrac32\langle\sigma_h^2\rangle$ is that of halo centre-of-mass motions. It is carried by the coherent flow and grows as in linear theory, $\langle K_{ph}\rangle\propto a$. The virial energy $\langle K_{pv}\rangle=\tfrac32\langle u^2\rangle$ is that of random motions inside halos, created by shell crossing and collapse. It is a purely nonlinear quantity and grows as $\langle K_{pv}\rangle=|\varepsilon_u|\,t$, where $\varepsilon_u$ is the constant rate of the inverse kinetic-energy cascade \citep{Xu:2021-Inverse-mass-cascade-mass-function}. Before $a_*$, most of the kinetic energy is coherent and the dynamics are linear; after $a_*$, most is random and the dynamics are nonlinear. The condition $\langle K_{pv}\rangle=\langle K_{ph}\rangle$ thus marks the transition from linear to nonlinear dynamics. Since $\langle K_{pv}\rangle/\langle K_{ph}\rangle\sim (M^*/m_h^*)^{2/3}$, it is equivalent to $M^*\simeq m_h^*$, up to factors of order unity. Geometrically, it corresponds to $R_L(M^*)\sim\Psi_\parallel$, where $R_L$ is the Lagrangian radius of mass $M^*$ and $\Psi_\parallel$ the rms displacement of the coherent linear flow: multistream regions come to span the coherent displacement.

\paragraph*{Total halo mass and the slope $\lambda_1$.}
After $a_*$, the mass flux below the cutoff is scale independent, and the halo group mass $m_g\propto m_h^{-\lambda_1}$, the total mass in halos of a given mass, is time invariant. This is small-scale permanence \citep{Xu:2023-Dark-matter-halo-mass-functions-and}. In the diffusion variable $y=m_h^{1-\lambda_1}/(1-\lambda_1)$, with $dy=m_h^{-\lambda_1}dm_h$, permanence means equal halo mass per unit $y$. The total halo mass is therefore the extent of the filled region in $y$, i.e.\ the position of the diffusion front, which grows as $\sqrt{C}$:
\begin{equation}
\label{eq:ts_three}
M_h\propto\int_0^{M^*}m_h^{-\lambda_1}\,dm_h\propto M^{*\,1-\lambda_1}\propto\sqrt{C}.
\end{equation}
After crossover $a_*$, both characteristic masses are such fronts on the same cascade clock $C=a$ but with different exponents: $m_h^*$ with $\lambda_2=2/3$, and $M^*$ with $\lambda_1$ (Eq.~\eqref{eq:3-18}). For $M^*$ this is the same front as above, rewritten on the cascade clock. Hence
\begin{equation}
\label{eq:ts_three_b}
M_h\propto M^{*\,1-\lambda_1}\propto m_h^{*\,1-\lambda_2}\propto a^{1/2}.
\end{equation}
The total halo mass thus grows diffusively, $M_h\propto a^{1/2}$, from $m_h^*\propto a^{3/2}$. With $M^*\propto a^3$, Eq.~\eqref{eq:ts_three_b} fixes $\lambda_1=5/6$, close to the fitted 0.878. This value of $\lambda_1$ assumes equal densities for all halos below the cutoff; the small excess of the fitted slope reflects the denser, early-formed small halos that retain the primordial imprint.

\begin{table}
\centering
\caption{The two characteristic masses before and after the crossover $a_*$. Exponents of $a$ are for matter domination.}
\label{tab:two_scales}
\small
\begin{tabular}{lcc}
\hline
Quantity & $a<a_*$ & $a>a_*$ \\
\hline
Ordering & $M^*<m_h^*$ & $m_h^*<M^*$ \\
$\lambda$ of $m_h^*$ & $2/3$ & $2/3$ \\
Clock of $m_h^*$ & $a^2$ & $a$ \\
$m_h^*$ & $a^3$ & $a^{3/2}$ \\
Clock of $M^*$ & $a^2$ & $a^2$ \\
$n$ at $M^*$ & $-7/5$ & $-1$ \\
$M^*$ & $a^{15/4}$ & $a^3$ \\
$M^*/m_h^*$ & $a^{3/4}$ & $a^{3/2}$ \\
$M_h$ & faster than $a^{1/2}$ & $M^{*\,1-\lambda_1}$ \\
Mass flux $\Pi_m$ & scale dependent & $\varepsilon_m=\dot M_h$ \\
Dominant energy & $\langle K_{ph}\rangle\propto a$ & $\langle K_{pv}\rangle\propto t$ \\
$\langle K_{pv}\rangle$ & \multicolumn{2}{c}{$|\varepsilon_u|\,t$} \\
\hline
\end{tabular}
\end{table}

Therefore, the model complements EPS in three ways: 1) a mass-dependent diffusivity from the merger waiting time $\tau_g$;
2) a double-$\lambda$ generalization that captures the transition from primordial to gravity-dominated dynamics through two clocks; and 3) a single Fokker--Planck language that connects the halo mass function, the density profile and turbulence.

\section{Particle random walk for halo structure}
\label{sec:4}
Similarly to the halo mass function, the internal structure of haloes, i.e. the spatial distribution of dark matter particles, can be described within the same stochastic framework. The halo density profile is also shaped jointly by primordial spectrum and gravitational dynamics. A steeper primordial spectrum (more negative index) generally produces haloes with lower concentrations and less dense central cusps. Because halo assembly proceeds inside-out, the innermost regions collapse earlier from high-redshift peaks in the initial density field; therefore, their structure retains a strong memory of the primordial spectrum. In contrast, the outer halo regions are assembled later through numerous successive mergers and smooth accretion events, which progressively erase the memory of the initial spectrum. Consequently, the dynamics near the characteristic scale $r_s$, and for haloes with a characteristic mass $m_h^*$, become less sensitive to primordial spectrum and are governed primarily by gravity, with only a weak dependence on the initial conditions. 

To model the halo density profile, we adopt two simplifying assumptions: accretion is smooth, and once bound, a particle remains fixed at its accretion radius. Each particle, therefore, contributes a single step to the random walk at position $\boldsymbol{X}$, deposited at the halo boundary and remains bound thereafter. Since every particle's step is drawn from the same underlying stochastic law, the ensemble of one-step particles is statistically equivalent to a single tracer executing many successive steps of that same continuous-time random walk, within a self-consistent mean gravitational potential set by the instantaneous particle distribution. This equivalence reduces the many-particle accretion problem to an effective single-particle random walk with position-dependent waiting time $\tau_{gr}(r)$, governed by a three-dimensional Fokker-Planck equation whose solution $P_r(r,a)$ yields the halo density profile $\rho_r(r,a)$ directly.

Similar to the halo mass function case, the scale factor $a$ serves as the natural clock for particle diffusion within haloes; it plays the role of the cascade clock $C=a$ of Section~\ref{sec:3}, appropriate for the gravity-dominated outer regions. The stochastic motion of an individual particle with position $\boldsymbol{X}$ is described by the Langevin equation
\begin{equation} 
\label{eq:4-1} 
\frac{d\boldsymbol{X} }{da} = \sqrt{2 D_P(\boldsymbol{X})} \boldsymbol{\xi}\left(a\right) \propto \frac{\delta x}{\tau_{gr}Ha}\propto \frac{\delta x}{r^{-\gamma}}.       
\end{equation} 
We assume a power-law form $\tau_{gr}(r)\propto r^{-\gamma}$, with $r\equiv |\boldsymbol{X}|$ the distance from the halo center, where the exponent $\gamma$ encodes the influence of the primordial spectrum and/or gravitational dynamics. The corresponding scale-dependent diffusivity reads 
\begin{equation} 
\label{eq:4-2} 
D_{P}(\boldsymbol{X})=D_{0} r^{2\gamma},         
\end{equation} 
where $D_{0}$ is a proportional constant. Thus diffusion is suppressed (and the waiting time enhanced) toward the halo center, leading to an increased particle density. In spherical symmetry, the associated Fokker-Planck equation for the radial distribution $P_r(r,a)$ is
\begin{equation} 
\label{eq:4-4} 
\frac{\partial P_{r}}{\partial a} = \left[\frac{\partial }{\partial r}+\frac{2}{r}\right]\left[ \alpha\frac{\partial D_P}{\partial r}P_{r} +D_P \frac{\partial P_r}{\partial r}\right], 
\end{equation} 
where $\alpha$ is the interpretation parameter. For the stochastic integral, the It$\hat{\textrm{o}}$ interpretation yields the limit of a discrete random process as the length of the discrete intervals tends towards zero. For the It$\hat{\textrm{o}}$ convention ($\alpha$=1; e.g.\ \citep{Sokolov:2010-Ito-Stratonovich-Hänggi}), which is physically appropriate for a discrete-jump process where the scattering event is evaluated at the pre-jump position, exact solution reads (Appendix~\ref{sec:app})
\begin{equation}
\label{eq:4-5} 
P_{r} \left(r,a\right) = \frac{(2-2\gamma)^{\frac{\gamma-2}{1-\gamma}} r^{-2\gamma}}{4\pi\left(D_0 a\right)^{\frac{3-2\gamma}{2-2\gamma}}\Gamma\left(\frac{3-2\gamma}{2-2\gamma}\right)}\exp \left(-\frac{r^{2-2\gamma}}{4(1-\gamma)^2D_0 a}\right).        
\end{equation} 
Identifying $P_r(r,a)$ with the halo density profile $\rho_r(r,a)$ yields $\rho_r\propto r^{-2\gamma}$ for a waiting time $\tau_{gr}(r)\propto r^{-\gamma}$.

\added{We emphasize that the quantity $\Phi_r(r)\equiv Gm_r(r)/r\propto v_c^2(r)$ used below to close the model is \emph{not} the true gravitational potential $\Phi(r)$ of the halo, which is negative and deepest at the center; rather, $\Phi_r(r)$ is a local quantity set by the enclosed mass and equivalent to the circular velocity squared, $v_c^2(r)$. In the gravity-dominated regime, and by analogy with the halo mass function case (Eq.~\eqref{eq:tau_phi}), we \emph{posit as a closure assumption} that the waiting time scales inversely with this quantity, $\tau_{gr}(r)\propto v_c^2(r)^{-1}\propto\Phi_r^{-1}$: where $v_c^2(r)$ is locally larger, particles complete an orbital/scattering cycle faster, shortening the residence time. This is a physically motivated \emph{closure ansatz}, like eddy turnover time for turbulence, rather than a first-principles derivation. In the gravity-dominated regime, similar to the halo mass function case (Eq. \eqref{eq:tau_phi}), the waiting time is set by the local potential}, 
\begin{equation}
\label{eq:4-5-2} 
\tau_{gr}\propto \Phi_r(r)^{-1}. 
\end{equation}

Next, for a power law waitint time $\tau_{gr}\propto r^{-\gamma}$, Fokker-Planck solution (Eq. \eqref{eq:4-5}) implies $\rho_r \propto r^{-2\gamma}$, so that the enclosed mass scales as $m_r \propto \rho_r r^3 \propto r^{3-2\gamma}$ and the local potential as $\Phi_r(r) \propto Gm_r/r \propto r^{2-2\gamma}$. The waiting time thus also satisfies $\tau_{gr}\propto \Phi_r^{-1}\propto r^{2\gamma-2}$ from the closure above. Requiring consistency with $\tau_{gr} \propto r^{-\gamma}$ yields
\begin{equation}
\label{eq:4-6} 
2\gamma-2=-\gamma \quad \textrm{such that}\quad \gamma=2/3, 
\end{equation} 
so that the asymptotic density slope is $2\gamma=4/3$, i.e. $\rho_r\propto r^{-4/3}$ and $\tau_{gr}\propto r^{-2/3}$. In the absence of any imprint from the initial spectrum, gravity alone therefore drives haloes toward a limiting slope of -4/3. For a realistic CDM spectrum, the core structure can be modified by primordial fluctuations, but slope -4/3 is expected to hold for massive haloes of characteristic mass $m_h^*$ and on scales near $r_s$. Around $r_s$, migration is gravity-dominated, giving $\gamma=2/3$, whereas on scales where the primordial spectrum dominates, $\gamma$ is dependent on the effective spectral index of the initial spectrum. 

\added{This gravity-dominated closure, and its breakdown near the halo center, parallels an independent statistical-mechanics argument by Pontzen and Governato \citep{Pontzen:2013-Conserved-actions-maximum-entropy}, who derive the halo phase-space distribution from a maximum-entropy argument constrained by conserved orbital actions rather than energy alone; their result matches $N$-body haloes closely except in the innermost cusp, which they attribute to the local orbital time-scale becoming too short for their constraint to remain valid. The present model additionally predicts a universal gravity-dominated slope -4/3, which holds around $r_s$ where mergers and phase mixing have established a quasi-steady state, but similarly breaks down at small $r$, where the profile instead retains the imprint of the primordial spectrum.}

High-resolution $N$-body simulations consistently produce cuspy halo density profiles that are shallower than isothermal at small radii and steeper at large radii \citep{Navarro:1997-A-universal-density-profile-fr, Navarro:2004-The-inner-structure-of-ACDM-ha}. Despite broad agreement on this qualitative shape, the precise value of the asymptotic inner slope $s$ remains unsettled: reported values range from $s>-1.0$ \citep{Navarro:2010-The-diversity-and-similarity-of-simulated} to $s=-1.2$ \citep{Diemand:2011-The-Structure-and-Evolution-of-Cold-Dark} and $s\approx -1.3$ \citep{Governato:2010-Bulgeless-dwarf-galaxies-and-dark-matter-cores, McKeown:2022-Amplified-J-factors-in-the-Galactic-Centre, Lazar:2020-A-dark-matter-profile-to-model-diverse}, and no first-principles theory uniquely predicts it \citep{Cooray:2002-Halo-models-of-large-scale-str}. The present paper offers a new perspective on this problem. Analogously to the halo random walk in mass space that produces the halo mass function (Section \ref{sec:3}), we model the redistribution of dark matter particles as a random walk in physical space (Section \ref{sec:4}). This framework yields two physically distinct regimes for the density profile. At small radii ($r\ll r_s$), the inner density slope $s$ is sensitive to the primordial power spectrum ($s\simeq -(3+n)$), reflecting the inside-out collapse of high-redshift density peaks; for a realistic CDM spectrum, the slope varies continuously with $r$ and has no universal asymptotic value. Near the characteristic scale radius $r_s$, however, repeated mergers, smooth accretion, violent relaxation, and phase mixing progressively erase the memory of the initial spectrum, and the density profile is governed primarily by nonlinear gravitational dynamics $s\simeq -4/3$. In this gravity-dominated transition regime, the model predicts a universal logarithmic slope of $-4/3$.

\begin{figure}
\includegraphics[width=\columnwidth]{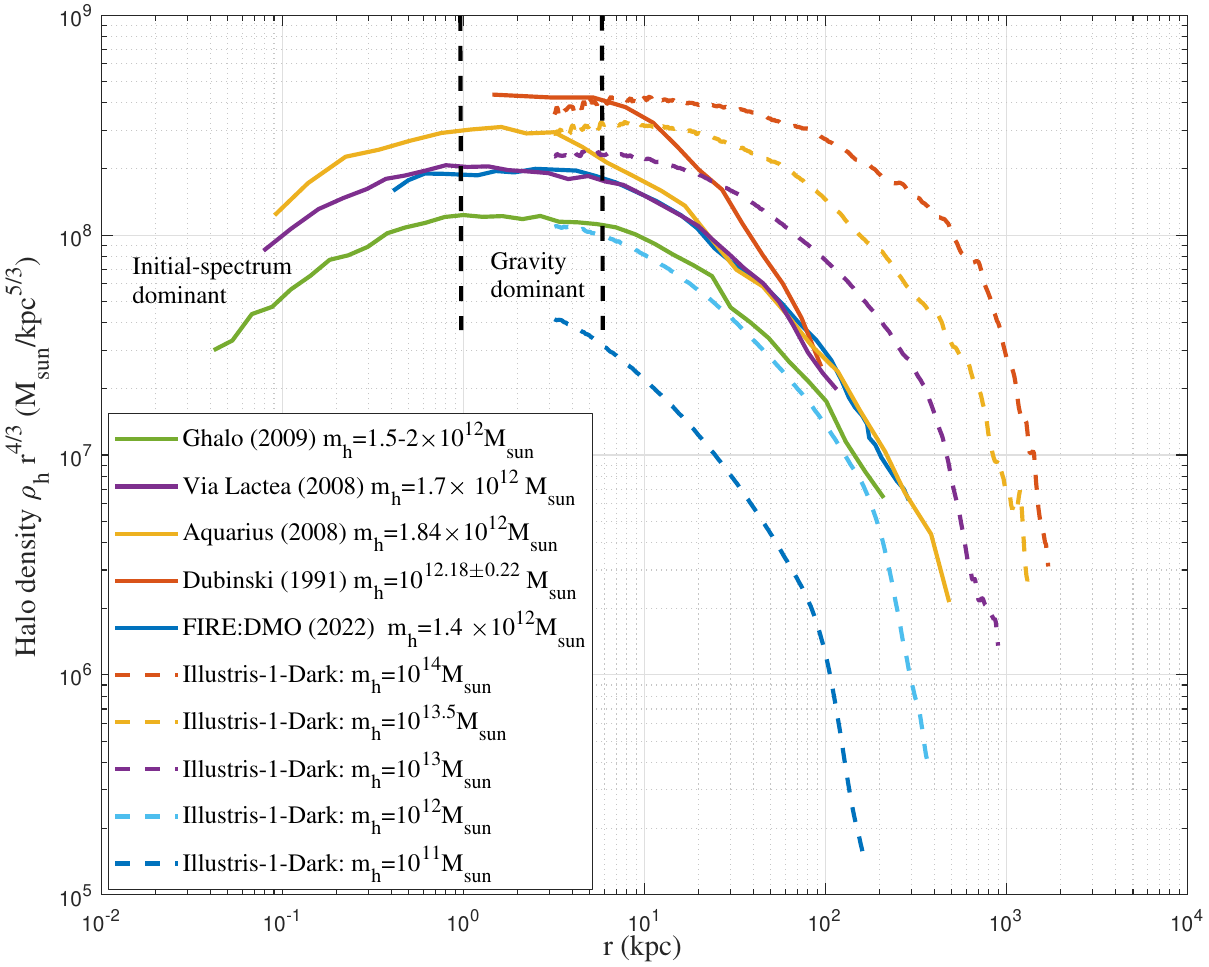}
\caption{Compensated halo density profiles $\rho_r(r)r^{4/3}$ for six sets of high-resolution $N$-body simulations: (1) Ghalo \citep{Stadel:2009-Quantifying-the-heart-of-darkness}; (2)~Via Lactea \citep{Diemand:2008-Clumps-and-streams-in-the-local}; (3)~Aquarius \citep{Springel:2008-The-Aquarius-Project}; (4)~Dubinski \citep{Dubinski:1991-The-Structure-of-Cold-Dark}; (5)~FIRE:DMO \citep{McKeown:2022-Amplified-J-factors-in-the-Galactic-Centre};
(6)~Illustris-1-Dark \citep{Vogelsberger:2014-Introducing-the-Illustris-Project}. Multiplying by $r^{4/3}$ removes the $-4/3$ slope, so a horizontal plateau in this representation indicates a density profile consistent with $\rho_r\propto r^{-4/3}$ (Eq. \eqref{eq:4-6}) dominated by gravitational dynamics. Most simulated haloes of $10^{12}M_{sun}$ show an approximately flat plateau on scales of several kpc (between the two dashed lines), whose extent grows with halo mass. On sub-kpc scales, the compensated profile declines at small $r$, indicating a shallower inner slope that is sensitive to the primordial spectrum.}
\label{fig:5}
\end{figure}

\begin{figure}
\includegraphics[width=\columnwidth]{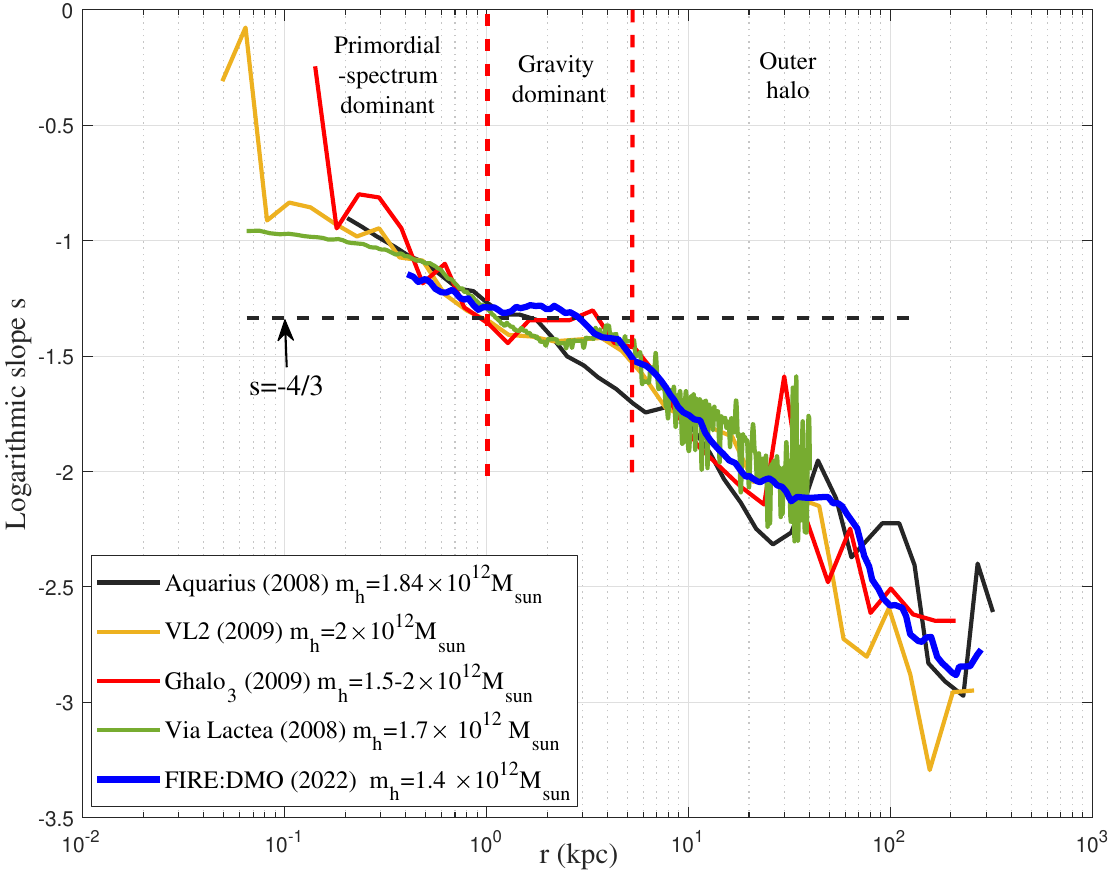}
\caption{Logarithmic density slope $s = d\ln\rho_r/d\ln r$ as a function of radius $r$, shown for five simulated dark matter haloes with masses around $10^{12}\,M_\odot$: Aquarius \citep{Springel:2008-The-Aquarius-Project}, VL2 and Ghalo$_3$ \citep{Stadel:2009-Quantifying-the-heart-of-darkness}, Via Lactea \citep{Diemand:2008-Clumps-and-streams-in-the-local}, and FIRE:DMO \citep{McKeown:2022-Amplified-J-factors-in-the-Galactic-Centre}. Three regimes are marked: a primordial-spectrum-dominated regime at small $r$, a gravity-dominated regime at intermediate $r$ (between the dashed red lines), and a linear regime at large $r$. Within the gravity-dominated regime, all profiles flatten to a common slope of approximately $s \approx -4/3$ (dashed black line) over the scale range $r \sim 1$ to 6 kpc, reflecting a universal density scaling once gravitational collapse dominates.}
\label{fig:9999-8888}
\end{figure}

Figures \ref{fig:5} and \ref{fig:9999-8888} demonstrates this prediction using compensated profiles $\rho_r\times r^{4/3}$ and logarithmic density slope $s=d\ln\rho_r/d\ln r$ for six independent simulation suites spanning more than three decades in halo mass. A horizontal plateau in this representation directly identifies a $-4/3$ density slope. These simulations show a clear plateau on scales of several kpc, with the plateau extending to larger radii for more massive haloes, consistent with the predicted gravity-dominated transition regime. At sub-kpc scales, the compensated profiles decrease toward smaller $r$, indicating shallower, spectrum-dependent inner slopes with no universal asymptotic value, also consistent with the primordial-spectrum-dominated inner regime of the model. The analytical slope -4/3 is in good agreement with these simulation datasets across the gravity-dominated regime.

Analogously to the halo mass function in Eq.~\eqref{eq:3-18}, the effective exponent $\gamma$ can take different values in two regimes: an inner region where the profile is sensitive to the primordial spectrum, and an outer region around $r_s$, where gravitational dynamics dominate. Adopting two exponents, $\gamma_1$ and $\gamma_2$, for these regimes, the corresponding double-$\gamma$ distribution, obtained from the single-$\gamma$ solution in Eq.~\eqref{eq:4-5}, is
\begin{equation}
\label{eq:4-7} 
P_{r} \left(r\right) = \frac{(2-2\gamma_2)^{\frac{2\gamma_1-2-\gamma_2}{1-\gamma_2}} r^{-2\gamma_1} }{4\pi\left(D_0 a\right)^{\frac{3-2\gamma_1}{2-2\gamma_2}}\Gamma\left(\frac{3-2\gamma_1}{2-2\gamma_2}\right)} \exp \left(-\frac{r^{2-2\gamma_2}}{4(1-\gamma_2)^2D_0 a}\right).
\end{equation} 
Introducing the conventional scale radius $r_s(a)$ where the logarithmic slope of $P_{r}(r,a)$ is equal to -2, we should have
\begin{equation}
\label{eq:4-8} 
4(1-\gamma_2)^2D_0 a = \frac{2-2\gamma_2}{2-2\gamma_1}r_s^{2-2\gamma_2}.         
\end{equation} 
Introducing a dimensionless spatial-temporal variable $x\equiv r/r_s(a)$, the resulting two-parameter radial distribution takes the form
\begin{equation}
\label{eq:4-9} 
P_{r} \left(x\right) = \frac{\alpha \beta^{-(\frac{1}{\alpha}+\frac{1}{\beta})}}{4\pi\Gamma\left(\frac{1}{\alpha}+\frac{1}{\beta}\right)} x^{\frac{\alpha}{\beta}-2} \exp\left(-\frac{x^{\alpha}}{\beta}\right),      
\end{equation} 
where two dimensionless parameters 
\begin{equation}
\label{eq:4-10} 
\alpha = 2-2\gamma_2 \quad \textrm{and} \quad \beta = \frac{1-\gamma_2}{1-\gamma_1}
\end{equation} 
are directly related to the exponents $\gamma_1$ and $\gamma_2$ that govern the scale dependence of the waiting time $\tau_{gr}(r)\propto r^{-\gamma}$. The double-$\gamma$ distribution in Eq.~\eqref{eq:4-9} reduces to the Einasto profile in Eq.~\eqref{eq:1-3} when $\alpha=2\beta$, corresponding to $\gamma_1=0$, or equivalently to an initial spectrum index $n=-3$ using the approximate relation $2\gamma=3+n$ \citep{Syer:1998-Dark-halo-mergers-and-the-formation, Xu:2021-Inverse-mass-cascade-mass-function}, where $2\gamma$ is the density slope. In the gravity-dominated regime, $\alpha=2\beta/3$ corresponds to $\gamma_1=2/3$.

The double-$\gamma$ form (Eq. \eqref{eq:4-7}) is not a free two-parameter ansatz but is the minimal smooth interpolation between two single-$\gamma$ regimes, as can be verified by examining its limiting behavior. For $r\ll r_s$, $P_r\propto r^{-2\gamma_1}$, recovering the inner single-$\gamma$ solution with exponent $\gamma_1$. For $r\sim r_s$, the exponential cutoff $\exp(-x^\alpha/\beta)$ in Eq. \eqref{eq:4-9} becomes significant and is governed entirely by $\gamma_2$ through $\alpha=2-2\gamma_2$, recovering the outer single-$\gamma$ behavior. The condition that the logarithmic slope equals $-2$ at $r=r_s$ (Eq. \eqref{eq:4-8}) uniquely fixes the matching between the two regimes, leaving no free parameters beyond $\gamma_1$ and $\gamma_2$ themselves, both of which are determined by the physical conditions (primordial spectrum and gravitational dynamics, respectively).

\section{Conclusions}
We have developed a unified stochastic random-walk and Fokker-Planck framework that describes the statistical evolution of three a priori distinct nonequilibrium systems, turbulent energy cascade \added{, cosmic dark matter halo assembly, and the internal structure of individual haloes}, within a common mathematical language. In all cases, the central object is a probability distribution in an appropriate scale space, whose evolution is governed by a Fokker-Planck equation with scale-dependent diffusivity arising from a scale-dependent waiting time between stochastic transfer events.

\added{For turbulence, modeling the energy cascade as a random walk of ``energy carriers'' in wavenumber space, with a power-law waiting time $\tau_{gk}\propto k^{-\gamma}$ and a $k^2$ viscous sink, yields the exact stationary spectrum of Eq.~\eqref{eq:2-6}: a power law $E(k)\propto k^{-(1+\gamma)}$ smoothly joined to an exponential dissipation-range cutoff. A self-consistency closure relating the waiting time to the local eddy-turnover time fixes $\gamma=2/3$, recovering the Kolmogorov $-5/3$ law together with the empirical Pao spectrum (Eq.~\eqref{eq:1-1}), and predicts the Kolmogorov constant $C_K=(\beta C_X)^{-2/3}/3$ in terms of the measurable cascade efficiency $C_X$, in good agreement with DNS and grid-turbulence measurements. Importantly, the scale-independence of the mean energy flux $\varepsilon$ in the inertial range, normally taken as an input hypothesis in the classical K41 argument, emerges here as a consequence of the stationary Fokker-Planck solution.}

For dark matter halos, this framework yields a Langevin equation for halo mass growth in which a clock set by the scale factor emerges as the natural time variable: the linear clock $C=a^2$ before the crossover $a_*$ and the cascade clock $C=a$ after. The resulting mass function reduces exactly to the Press-Schechter form in the linear-theory limit, but generalizes naturally to a double-$\lambda$ form that captures the transition from the primordial-spectrum-dominated regime at small masses to the gravity-dominated regime near the characteristic mass $m_h^*$. An analogous random walk of dark matter particles in physical space, with waiting time $\tau_{gr}\propto r^{-\gamma}$, yields a double-$\gamma$ density profile that reduces to the Einasto form and predicts a gravity-dominated asymptotic slope of $-4/3$ near the scale radius $r_s$, while the inner slope retains a direct imprint of the primordial spectral index. The framework is not limited to these three applications: even the baryonic density profile in galaxies follows similar scaling laws \citep{Xu:2024-Cosmic-quenching-and-scaling-laws}. More generally, any system in which energy, mass, or another conserved quantity cascades hierarchically through scale space, with the local transfer rate set by the dynamics at each scale, admits a Fokker-Planck description of the same form. This suggests broad applicability across astrophysics and beyond, to any setting where emergent spectra or mass distributions arise from scale-by-scale stochastic dynamics \citep{Xu:2026-Nonequilibrium-training-dynamics}. A rigorous mathematical connection between this stochastic formulation and the underlying deterministic equations, the Navier–Stokes equation and the collisionless Boltzmann–Vlasov equation, remains an open problem, left for future work.

\section*{Author contributions}
Z.J. Xu performed the conception, data analysis, and writing.

\section*{Funding}
This research was supported by the Laboratory Directed Research and Development at Pacific Northwest National Laboratory (PNNL). PNNL is a multiprogram national laboratory operated for the U.S. Department of Energy (DOE) by Battelle Memorial Institute under contract no. DE-AC05-76RL01830.

\section*{Data availability}
The datasets analysed in this study are all publicly available and were previously published: Illustris and IllustrisTNG data are available at https://www.illustris-project.org and https://www.tng-project.org \citep{Vogelsberger:2014-Introducing-the-Illustris-Project, Springel:2018-First-results-from-the-IllustrisTNG}; the Virgo SCDM simulation data are described in \citep{Jenkins:1998-Evolution-of-structure-in-cold}; and the Ghalo, Via Lactea, Aquarius, Dubinski, and FIRE:DMO halo density profiles are as published in \citep{Stadel:2009-Quantifying-the-heart-of-darkness}, \citep{Diemand:2008-Clumps-and-streams-in-the-local}, \citep{Springel:2008-The-Aquarius-Project}, \citep{Dubinski:1991-The-Structure-of-Cold-Dark}, and \citep{McKeown:2022-Amplified-J-factors-in-the-Galactic-Centre}, respectively. The data extracted from published simulations are available from the corresponding author upon reasonable request.

\section*{Code availability}
This study is theoretical and analytical; all results were obtained by closed-form solutions presented in the text. No specialized simulation code was developed. Standard curve-fitting and plotting scripts are available at \href{https://doi.org/10.5281/zenodo.22051522}{10.5281/zenodo.22051522}.

\appendix
\addtocontents{toc}{\protect\setcounter{tocdepth}{-1}}

\section{Exact solution of general Fokker-Planck equation}
\label{sec:app}
In this section, we derive exact solutions of a general $d$-dimensional Fokker-Planck equation in scale space $x$, with a power-law diffusivity $D_P(x)$ and a possible dissipation term. The equation reads
\begin{equation} 
\label{eq:a-1} 
\begin{split}
&\frac{\partial P}{\partial t} = \left(\frac{\partial }{\partial x}+\frac{d-1}{x}\right)\left[ \alpha\frac{\partial D_P}{\partial x}P +D_P \frac{\partial P}{\partial x}\right]-2\nu x^2P, \\
&D_P(x) = D_0x^{2\gamma}=D_0x^g,
\end{split}
\end{equation} 
where $d$ is the dimension, $\alpha$ is the interpretation parameter, $\nu$ is the viscous dissipation. The associated waiting time obeys $\tau_g(x)\propto x^{-\gamma}$ with $g=2\gamma$ (Eq. \eqref{eq:2-3}). We look for a similarity solution of
\begin{equation} 
\label{eq:a-2} 
P(x,t)=\frac{A}{\left(D_0t\right)^bx^a}\exp\left[-\frac{x^c}{f\left(D_0t\right)^h}\right],
\end{equation} 
where the normalization constant A is fixed by $\int_0^{\infty}P(x)dx=1$. Substituting Eq.~\eqref{eq:a-2} into Eq.~\eqref{eq:a-1} yields the algebraic constraint
\begin{equation} 
\label{eq:a-3} 
\begin{split}
&f^2(D_0t)^{2h+1}x^{g}(a-\alpha g)(a-g+2-d)\\
&+fc(D_0t)^{h+1}x^{c+g}(c+d-2-2a+g+\alpha g)\\
&-bf^2(D_0t)^{2h}x^2+hf(D_0t)^hx^{c+2}\\
&-c^2(D_0t)x^{2c+g}+2f^2(\nu/D_0)(D_0t)^{2h+1}x^4=0.
\end{split}
\end{equation} 
For three cases, the parameters must take the following forms:
\begin{enumerate}
\item \noindent Stationary turbulence spectrum ($d=1$, $h=b=0$, and $c\neq 0$):
\begin{equation} 
\label{eq:a-4} 
a=1+\gamma,\quad c=2-\gamma,\quad \alpha=\frac{1}{2}\left(1+\frac{1}{\gamma}\right),\quad f=(2-\gamma)\sqrt{\frac{D_0}{2\nu}}.
\end{equation} 
To prove that $\alpha=(1+1/\gamma)/2$ is the \emph{unique} admissible value, we examine the algebraic constraint Eq. \eqref{eq:a-3} for the stationary case. First, $2c+g=4$ such that the last two terms in Eq. \eqref{eq:a-3} can be canceled. The coefficient of $x^g$ must vanish identically. This gives two roots: either $a=\alpha g=2\alpha\gamma$, or $a=g-1=2\gamma-1$. \added{The second root, $a=2\gamma-1$ fixes the power-law exponent of $P(x)$ to $a=2\gamma-1$ independently of $\alpha$. For $\gamma=2/3$ this branch gives $P(k)\propto k^{-1/3}$, which does not reproduce the observed inertial-range scaling $-5/3$ for any choice of $\alpha$; this branch is therefore an admissible solution of the Fokker-Planck equation, but not the one describing the Kolmogorov cascade.} For the physical inertial-range scaling $P(k)\propto k^{-(1+\gamma)}$ we need $a=1+\gamma$. Therefore the first root must hold: $1+\gamma=2\alpha\gamma$, giving $\alpha=(1+\gamma)/(2\gamma)=(1+1/\gamma)/2$. For $\gamma=2/3$ this evaluates to $\alpha=5/4$. \added{Only this first branch simultaneously matches the required exponent \emph{and} uniquely fixes $\alpha$; hence it is selected on physical grounds (reproducing the correct inertial-range power law), while its associated value of $\alpha$ is then} a mathematical necessity, not a free choice. Appendix \ref{sec:appen2} provides more insights.
\\
\item \noindent Halo mass function ($d=1$, $\nu=0$, and $\alpha=1/2$):
\begin{equation} 
\label{eq:a-5} 
a=2\alpha\gamma,\quad c=2-2\gamma,\quad f=4(1-\gamma)^2, h=1, b=\frac{1-2\alpha\gamma}{2(1-\gamma)}.
\end{equation} 
The Stratonovich convention $\alpha=1/2$ is selected here because halo mergers arise as the continuum limit of a correlated (smooth) process with finite merger timescale, for which the Wong-Zakai theorem guarantees convergence to the Stratonovich interpretation. Although each accretion step in Fig.~\ref{fig:S2} is discrete, a halo undergoes many correlated encounters within one merger timescale, so the coarse-grained mass growth is a smooth process with finite correlation time. In contrast, each particle in Section~\ref{sec:4} makes a single jump, evaluated at its pre-jump position, which selects the It\^{o} convention.
\\
\item \noindent Halo density profile ($d=3$, $\nu=0$, and $\alpha=1$):
\begin{equation} 
\label{eq:a-6} 
a=2\alpha\gamma,\quad c=2-2\gamma,\quad f=4(1-\gamma)^2, h=1, b=\frac{3-2\alpha\gamma}{2(1-\gamma)}.
\end{equation} 
The It\^{o} convention $\alpha=1$ is appropriate here because it yields the limit of a discrete random process as the length of the discrete intervals tends towards zero \citep{Sokolov:2010-Ito-Stratonovich-Hänggi}. Dark matter particle scattering in a gravitational potential is a discrete-jump process: the transition probability is evaluated at the pre-jump position.
\end{enumerate}

\section{Effective inertial-range EDQNM closure}
\label{sec:appen2}
In this section, we show that the inertial-range Fokker-Planck model (Eq. \eqref{eq:2-5}) and Langevin equation in k-space (Eq. \eqref{eq:2-3}) and  can be derived from an effective inertial-range EDQNM closure. 

The standard Eddy Damped Quasi-Normal Markovian (EDQNM) transfer integral reads (\cite[Eq. (7.22)]{lesieur:2008-turbulence} \cite{Orszag:1970-Analytical-theories-of-turbulence}) 
\begin{equation}
T(k,t)=\iint_{\Delta_k}\!\Theta_{kpq}b_{kpq} E(q)\Big[k^2E(p)-p^2{E(k)}\Big]\frac{kdp\,dq}{pq}.
\label{eq:EDQNM}
\end{equation}
Here $\Delta_k$ is the triad domain,
\begin{equation}
\Delta_k=\{(p,q): |p-q|\le k\le p+q\},
\end{equation}
$\Theta_{kpq}$ is the Markovian triad memory time, $b_{kpq}$ is the geometrical interaction coefficient, and $E(k)$ is the one-dimensional isotropic energy spectrum. The gain-loss bracket in \eqref{eq:EDQNM} may be rewritten as
\begin{align}
k^2E(p)E(q)-p^2E(k)E(q) &= k^2p^2E(q) \left[ \frac{E(p)}{p^2}-\frac{E(k)}{k^2}\right].
\label{eq:EDQNM_balance_variable}
\end{align}
Thus the standard EDQNM transfer is driven by differences of the modal energy ${E(k)}/{k^2}$ (energy per mode). Consequently, the standard EDQNM/Leith local expansion has the reference spectrum $E_{\rm eq}(k)\propto k^2$. This is the absolute-equilibrium spectrum corresponding to equipartition of energy among Fourier modes.

In the present model, however, the energy-containing range $k<k_L$ is not resolved. The model is intended only for the inertial and dissipation ranges with integral scale $k_L\rightarrow0$. Therefore the infrared equilibrium $E_{\rm eq}\propto k^2$ is not part of the resolved dynamics. We instead introduce an inertial-range reference spectrum
\begin{equation}
E_{\rm ref}(k)\propto k^{-1-\gamma}>0,
\label{eq:Pref_def}
\end{equation}
a statically steady state spectrum. The purpose of this section is to demonstrate, at the level of an effective projected inertial-range closure, the replacement
\begin{equation}
\frac{E(k)}{k^2} \quad\longrightarrow\quad \frac{E(k)}{E_{\rm ref}(k)}
\end{equation}
leads to an effective inertial-range EDQNM closure and corresponding inertial-range Fokker-Planck equation and diffusion models.

We assume that after eliminating the energy-containing range $k<k_L$, the resolved inertial-range transfer can be represented by an effective gain-loss operator
\begin{equation}
T_{\rm eff}(k) = \int_{k_L}^{\infty} \left[W(k|p)E(p)-W(p|k)E(k)\right]\,dp.
\label{eq:effective_master}
\end{equation}
Here $W(k|p)$ is the effective transition rate from shell $p$ to shell $k$ in the resolved inertial range. We impose the following assumptions: 1) Positivity. $W(k|p)\ge 0$; 2) Locality. The kernel is concentrated near $p=k$: $p=k+\xi$ for $|\xi|\ll k$; 3) Inertial-range detailed balance. The projected kernel satisfies detailed balance with respect to the chosen inertial-range reference spectrum $E_{\rm ref}$: 
\begin{equation}
W(k|p)E_{\rm ref}(p) = W(p|k)E_{\rm ref}(k).
\label{eq:detailed_balance_Pref}
\end{equation}
This assumption means that $E_{\rm ref}$ is the reference measure of the projected inertial-range kernel. It is not the absolute equilibrium of the full Navier--Stokes system; rather, it is the effective cascade background after the energy-containing range range has been removed.

Using \eqref{eq:detailed_balance_Pref}, the gain term in \eqref{eq:effective_master} can be rewritten as 
\begin{align}
W(k|p)E(p) = W(k|p)E_{\rm ref}(p) &\frac{E(p)}{E_{\rm ref}(p)} \nonumber\\
&=W(p|k)E_{\rm ref}(k) \frac{E(p)}{E_{\rm ref}(p)}.
\end{align}
Similarly, the loss term is
\begin{equation}
W(p|k)E(k) = W(p|k)E_{\rm ref}(k) \frac{E(k)}{E_{\rm ref}(k)}.
\end{equation}
Therefore, the net difference term in Eq. \eqref{eq:effective_master} reads
\begin{align}
W(k|p)E(p)-&W(p|k)E(k)= \nonumber \\
& W(p|k)E_{\rm ref}(k) \left[\frac{E(p)}{E_{\rm ref}(p)} - \frac{E(k)}{E_{\rm ref}(k)} \right].
\end{align}
Substituting this into \eqref{eq:effective_master} gives
\begin{equation}
T_{\rm eff}(k) = \int_{k_L}^{\infty} \mathcal{K}(k,p) \left[\frac{E(p)}{E_{\rm ref}(p)} - \frac{E(k)}{E_{\rm ref}(k)} \right]\,dp,
\label{eq:Teff_ratio}
\end{equation}
where
\begin{equation}
\mathcal{K}(k,p):= W(p|k)E_{\rm ref}(k).
\label{eq:K_def}
\end{equation}
Equation \eqref{eq:Teff_ratio} proves that the variable
\begin{equation}
H(k):=\frac{E(k)}{E_{\rm ref}(k)}
\label{eq:H_def}
\end{equation}
is the natural detailed-balance variable of the inertial-range transfer. Let $p=k+\xi$ with $|\xi|\ll k$, and define $H(k)={E(k)}/{E_{\rm ref}(k)}$, Taylor expansion yields
\begin{equation}
H(p)-H(k) = H(k+\xi)-H(k) = \xi H_k+\frac{\xi^2}{2}H_{kk} +\mathcal{O}(\xi^3).
\label{eq:H_expansion}
\end{equation}
where $H_k$ representing the derivative with respect to $k$. Substituting \eqref{eq:H_expansion} into \eqref{eq:Teff_ratio} gives the second-order Kramers--Moyal expansion
\begin{equation}
T_{\rm eff}(k) = M_1(k)H_k + \frac{1}{2}M_2(k)H_{kk} + \mathcal{O}(\xi^3),
\label{eq:KM_H}
\end{equation}
with moments
\begin{equation}
M_n(k) = \int \xi^n\mathcal{K}(k,k+\xi)\,d\xi.
\label{eq:Mn_def}
\end{equation}
With $k_L\rightarrow0$, conservation of the nonlinear transfer in the resolved range, up to boundary fluxes at $k=k_L$ and $k=\infty$, requires that the local second-order expansion in divergence form. Thus
\begin{equation}
M_1(k)=\frac{1}{2}\frac{dM_2(k)}{dk}.
\label{eq:moment_relation}
\end{equation}
Define $D_{\rm eff}(k) := M_2(k)/2$, then \eqref{eq:KM_H} becomes
\begin{equation}
T_{\rm eff}(k) = \frac{\partial}{\partial k} \left[D_{\rm eff}(k)\frac{\partial H}{\partial k} \right] = \frac{\partial}{\partial k} \left[D_{\rm eff}(k) \frac{\partial}{\partial k} \left( \frac{E(k)}{E_{\rm ref}(k)} \right) \right]. 
\label{eq:Teff_H}
\end{equation}

It is convenient to define the Fokker-Planck diffusivity $D_P(k)$ by
\begin{equation}
D_{\rm eff}(k)=D_P(k)E_{\rm ref}(k).
\label{eq:DP_def}
\end{equation}
Therefore, effective inertial-range energy transfer function reads
\begin{equation}
T_{\rm eff}(k,t) = \frac{\partial}{\partial k} \left[ D_P(k)E_{\rm ref}(k) \frac{\partial}{\partial k} \left(\frac{E(k,t)}{E_{\rm ref}(k)} \right)\right].
\label{eq:generalized_Leith}
\end{equation}
Equation \eqref{eq:generalized_Leith} is the generalized Leith/Fokker-Planck operator associated with the projected inertial-range reference spectrum $E_{\rm ref}$. 

Expanding the bracket in \eqref{eq:generalized_Leith}, we find 
\begin{equation}
T_{\rm eff}(k) = \frac{\partial}{\partial k}\left[D_P(k)\frac{\partial E}{\partial k}-D_P(k)\left(\frac{d}{dk}\ln E_{\rm ref}(k)\right)\,E\right].
\label{eq:Teff_expanded}
\end{equation}
Now choose the inertial-range reference spectrum
\begin{equation}
E_{\rm ref}(k)=C_{\rm ref} k^{-1-\gamma}.
\label{eq:Pref_powerlaw}
\end{equation}
Then Substituting \eqref{eq:Pref_powerlaw} into
\eqref{eq:Teff_expanded} gives
\begin{equation}
T_{\rm eff}(k,t) = \frac{\partial}{\partial k} \left[D_P(k)\frac{\partial E(k,t)}{\partial k}+\frac{1+\gamma}{k}D_P(k)E(k,t)\right].
\label{eq:Teff_powerlaw}
\end{equation}
Assume that the effective inertial-range diffusivity scales as
\begin{equation}
D_P(k)=D_0 k^{2\gamma},
\label{eq:DP_powerlaw}
\end{equation}
Using \eqref{eq:DP_powerlaw}, the transfer function \eqref{eq:Teff_powerlaw} becomes
\begin{equation}
T_{\rm eff}(k,t) = \frac{\partial}{\partial k} \left[D_P(k)\frac{\partial E(k,t)}{\partial k}+\frac{1+\gamma}{2\gamma}\frac{dD_P}{dk}E(k,t)\right].
\label{eq:Teff_final}
\end{equation}

Including viscous dissipation, the final spectral equation is
\begin{equation}
\frac{\partial E}{\partial t}=\frac{\partial}{\partial k}\left[D_P(k)\frac{\partial E}{\partial k}+\alpha\frac{dD_P(k)}{dk}E\right]-2\nu k^2E,
\label{eq:FP_final}
\end{equation}
with $\alpha={1+\gamma}/({2\gamma})$, which is the inertial-range Fokker-Planck model (Eq. \eqref{eq:2-5}) we obtained for random walk in k-space.

The standard Leith model is recovered from \eqref{eq:generalized_Leith} by choosing
\begin{equation}
E_{\rm ref}(k)=k^2.
\end{equation}
Then we obtain
\begin{equation}
T(k)=\frac{\partial}{\partial k}\left[D_L(k)k^4\frac{\partial}{\partial k}\left(\frac{E}{k^2}\right)\right],
\end{equation}
which is exactly the usual Leith diffusion operator ($D_P(k)=k^2D_L(k)$). Thus, the proposed model is a direct generalization of the Leith operator in which the reference measure $k^2$ is replaced by the projected inertial-range reference spectrum $E_{\rm ref}$.

The derivation demonstrates the following conditional result:
\begin{quote}
If the projected inertial-range transfer can be represented as a local gain-loss process satisfying detailed balance with respect to a positive reference spectrum $E_{\rm ref}$, then the natural transported variable is $E/E_{\rm ref}$, and the local second-order expansion yields the generalized Leith operator \[T_{\rm eff}(k)=\partial_k\left[D_P(k)E_{\rm ref}(k)\partial_k\left(\frac{E}{E_{\rm ref}}\right)\right].\]
\end{quote}

However, the replacement ${E}/{k^2} \rightarrow {E}/{E_{\rm ref}}$ is not an identity at the level of the bare EDQNM integral derived directly from the Navier-Stokes equations, which is still an open question. The bare EDQNM bracket fixes the reference measure $k^2$. The modified model should therefore be interpreted as an effective projected inertial-range closure, in which the forcing range $k<k_L$ has been removed in the limit of $k_L\rightarrow0$.

\section{Comparison with the K41 theory}
\label{sec:comparison}
\subsection{K41 Theory: Assumptions and Dimensional Argument} 
The classical theory of turbulence \citep{Kolmogoroff:1941-Dissipation-of-energy-in-the-l}, commonly referred to as K41, rests on three central assumptions. 
\begin{enumerate}
\item \noindent \textit{Local isotropy and homogeneity.} At sufficiently high Reynolds numbers, the small-scale motions are assumed to be statistically homogeneous and isotropic, regardless of the large-scale anisotropy of the forcing.
\item \noindent \textit{Universal equilibrium of small scales.} The statistics of small-scale eddies (those in and beyond the inertial range) are assumed to be universal, i.e. independent of the large-scale geometry, boundary conditions, or the specific mechanism by which energy is injected.
\item \noindent \textit{Scale-independent mean energy dissipation rate.} In the inertial range $k_L \ll k \ll k_\eta$, energy is transferred conservatively across scales at a constant, scale-independent rate $\varepsilon$.  The energy spectrum is assumed to depend only on $\varepsilon$ and the wavenumber $k$; the kinematic viscosity $\nu$ is irrelevant in this range.
\end{enumerate}
Under these assumptions, a straightforward dimensional analysis using $[\varepsilon]=\mathrm{m^2\,s^{-3}}$ and $[k]=\mathrm{m^{-1}}$ yields the -5/3 power law 
\begin{equation}
E(k) = C_K \,\varepsilon^{2/3} k^{-5/3},
\label{eq:K41}
\end{equation}
where $C_K \approx 1.5$-$1.7$ is the Kolmogorov constant. The Kolmogorov microscale $\eta=(\nu^3/\varepsilon)^{1/4}$ provides the only scale at which viscous effects become $\mathcal{O}(1)$, and signals the onset of the dissipation range.
 
Although dimensionally compelling and empirically well-supported in the inertial range, the K41 framework has several important limitations.
\begin{itemize}
\item The spectrum in Eq.~\eqref{eq:K41} is valid \textit{only} in the inertial range.  Outside this range, viscous corrections must be added empirically; the widely used Pao spectrum \citep{Pao:1965-Structure-of-Turbulent-Velocity}
\begin{equation}
E(k)=C_K \varepsilon^{2/3}k^{-5/3}\exp\!\left[-\tfrac{3}{2}C_K(\eta k)^{4/3}\right]
\label{eq:Pao}
\end{equation}
appends an exponential dissipation-range cutoff by hand rather than deriving it from first principles.
\item The spectrum constant $C_K$ is not predicted by K41; it must be fitted to experimental data.
\item The assumption of a scale-independent $\varepsilon$ is taken as a given, not derived from the dynamics.
\item K41 does not address the full crossover region between the inertial and dissipation ranges.
\item Dimensional argument provides no natural extension to other nonequilibrium systems such as dark matter structure formation.
\end{itemize}
 
\subsection{Present Stochastic Framework: Assumptions\&Derivation}
The present work complments the K41 dimensional argument with an explicit dynamical model formulated as a random walk in wavenumber space. The key assumptions are as follows. 
\begin{enumerate}
\item \noindent \textit{Scale-locality of the cascade.} Triadic interactions are dominated by local modes with $|\mathbf{k}|\sim|\mathbf{p}|\sim|\mathbf{q}|$, so that energy is transferred between neighbouring shells in $k$-space.  This is the same scale-locality property invoked in K41 phenomenology, but here it is implemented explicitly as a nearest-neighbour random walk (Section~\ref{sec:2}).
\item \noindent \textit{Energy conservation in the absence of forcing and dissipation.} All triadic interactions conserve total energy. Energy is redistributed among "energy carriers" in $k$-space without any net creation or destruction, so that the stochastic dynamics merely drives diffusion.
\item \noindent \textit{Viscous dissipation as a scale-dependent sink.} Viscosity removes kinetic energy at a rate $2\nu k^2 E(k)$, entering the Fokker-Planck equation (Eq. \eqref{eq:2-5}) as a local $k$-dependent particle sink $-2\nu k^2 P(k)$.  No empirical bridge function is needed.
\item \noindent \textit{Power-law waiting time in $k$-space.} The waiting time between successive random-walk steps is assumed to satisfy $\tau_{gk}(k)\propto k^{-\gamma}$, parameterising the eddy turnover time. The exponent $\gamma$ is \textit{not} prescribed by dimensional analysis but is determined self-consistently.
\end{enumerate}
 
Under these assumptions, the energy cascade is described by the Langevin equation (Eq.~\eqref{eq:2-3}) and the corresponding Fokker-Planck equation (Eq.~\eqref{eq:2-5}) with a scale-dependent diffusivity $D_P(k)=D_0 k^{2\gamma}$.  The exact stationary solution (Eq.~\eqref{eq:2-6}) is
\begin{equation}
    P(k) \propto k^{-(1+\gamma)}\exp\!\left[-\lambda\!\left(\frac{k}{k_\eta}\right)^{2-\gamma}\right],
    \label{eq:Pk}
\end{equation}
which simultaneously gives the inertial-range power law $E(k)\propto k^{-(1+\gamma)}$, the exponential dissipation-range cutoff $\exp[-(k/k_\eta)^{2-\gamma}]$, and the normalisation constant $A$ (and hence $C_K$) from the constraint $\int P(k)\,dk=1$. Exponent $\gamma=2/3$ is determined self-consistently, which in turn recovers the K41 exponent $1+\gamma=5/3$.
 
\subsection{Key Differences and Contributions}
Table \ref{tab:comparison} summarizes the principal differences between K41 and the present framework.
\begin{table*}
\centering
\caption{Comparison between the classical K41 theory and the present stochastic random-walk framework.}
\label{tab:comparison}
\begin{tabular}{p{0.35\columnwidth} p{0.6\columnwidth} p{0.9\columnwidth}}
\hline
\textbf{Aspect} & \textbf{K41} & \textbf{Present work} \\
\hline
\makecell[l]{Method} 
    & Dimensional analysis using $\varepsilon$ and $k$ 
    & Fokker-Planck equation in $k$-space with scale-dependent diffusivity \\[4pt]
\hline
\makecell[l]{Inertial-range spectrum} 
    & $E(k)\propto \varepsilon^{2/3}k^{-5/3}$; exponent assumed by dimensional argument 
    & $E(k)\propto k^{-(1+\gamma)}$; exponent $\gamma=2/3$ determined self-consistently \\[4pt]
    \hline
\makecell[l]{Dissipation range} 
    & Not derived; empirical exponential cutoff appended post hoc (Pao spectrum) 
    & Exponential cutoff emerges naturally in the exact analytic solution \\[4pt]
    \hline
\makecell[l]{Spectrum constant $C_K$} 
    & Undetermined; fitted to experiments 
    & Fixed by the stationary distribution $P(k)$ \\[4pt]
    \hline
\makecell[l]{Energy flux $\varepsilon$} 
    & Prescribed as a constant input parameter; scale-independence assumed a priori 
    & Scale-independent $\varepsilon$ is a \textit{consequence} of the stationary solution, not a premise \\[4pt]
    \hline
\makecell[l]{Range of validity} 
    & Inertial range only 
    & Full spectrum from inertial to dissipation range with crossover \\[4pt]
    \hline
\makecell[l]{Role of viscosity} 
    & Sets the Kolmogorov scale $\eta$; not part of the spectral derivation 
    & Enters the Fokker-Planck equation as a $k$-dependent sink term \\[4pt]
    \hline
\makecell[l]{Unification} 
    & Restricted to hydrodynamic turbulence 
    & Same framework applied to halo mass function and density profile \\[4pt]
\hline
\end{tabular}
\label{table:C1}
\end{table*}
More specifically, the present framework advances beyond K41 in the following ways.
\paragraph{Self-consistent derivation of the -5/3 exponent.}
In K41, the $-5/3$ exponent is dictated purely by dimensional analysis once $\varepsilon$ is identified as the only relevant parameter. In the present work, the exponent $\gamma$ of the waiting time $\tau_{gk}(k)\propto k^{-\gamma}$ is a free parameter that is determined self-consistently (Eq. \eqref{eq:2-15}), yielding $\gamma=2/3$ and hence the $-5/3$ spectral slope.  The derivation, therefore, does not invoke dimensional analysis but instead identifies the spectral exponent as a dynamical fixed point of the Fokker--Planck equation.
 
\paragraph{First-principles derivation of the full spectrum.}
K41 produces only the inertial-range scaling law; bridging to the dissipation range requires an empirical choice (e.g. the Pao spectrum in Eq. \eqref{eq:Pao}). The exact stationary solution of the Fokker-Planck equation (Eq. \eqref{eq:Pk}) contains both the inertial-range power law and the exponential cutoff as a single analytic expression, describing the inertial-dissipation crossover without any empirical fitting.
 
\paragraph{Prediction of the spectrum constant.}
The normalization constraint $\int_0^\infty P(k)\,dk=1$, together with the solution in Eq.~\eqref{eq:Pk} and the definition $P(k)=2E(k)/u_0^2$, fixes the proportionality constant $A$ and therefore the Kolmogorov constant $C_K$ (Eq. \eqref{eq:2-18}). K41 does not predict $C_K$.
 
\paragraph{Scale-independent flux as a result, not an assumption.}
K41 assumes that the mean energy transfer rate $\varepsilon$ is constant and scale-independent throughout the inertial range; this is an input hypothesis.  In the random-walk picture, the stationarity of the Fokker--Planck solution implies a divergence-free probability current in $k$-space, which, when translated back to the energy language, corresponds to a constant energy flux.  Scale-independence of $\varepsilon$ thus emerges as a consequence of the stationary dynamics rather than being assumed.
 
\paragraph{Explicit role of the stochastic interpretation.}
K41 makes no reference to the mathematical convention used to integrate stochastic differential equations.  In the present framework, the Fokker-Planck equation explicitly contains an interpretation parameter $\alpha$, and the self-consistent stationary solution constrains $\alpha=(1+1/\gamma)/2$, providing a direct physical interpretation of the stochastic calculus appropriate for turbulence.
 
\paragraph{Unified framework across systems.}
Perhaps the most significant departure from K41 is the generality of the framework.  Whereas K41 is intrinsically tied to the Navier--Stokes cascade, the same random-walk Fokker-Planck construction, with a power-law waiting time in the appropriate "scale space", is applied in this paper to two other nonequilibrium systems. A random walk in halo mass space $m_h$ with waiting time $\tau_g\propto m_h^{-\lambda}$ yields the halo mass function (Section~\ref{sec:3}), reproducing the Press-Schechter form and predicting the double-$\lambda$ generalization without invoking spherical collapse.  A random walk in physical space with waiting time $\tau_{gr}\propto r^{-\gamma}$ yields the halo density profile (Section~\ref{sec:4}), reproducing the Einasto profile. In all three cases, the key spectral or structural parameters emerge self-consistently determined by the local dynamics, rather than as inputs to a dimensional argument.

\section{Characteristic mass \texorpdfstring{$\lowercase{{m_h}}^*$}{}}
\label{sec:chara_mass}
In this section, we derive the halo center-of-mass velocity dispersion $\sigma_h^2(m_h,a)$ from linear perturbation theory and the internal virial velocity dispersion $\sigma_v^2(m_h,a)$ from the virial theorem. The characteristic mass $m_h^*$ is fixed by the crossing condition 
\begin{equation}
\label{eq:crossing-def}
\sigma_h^2(m_h^*,a) = \sigma_v^2(m_h^*,a),
\end{equation}
marking the linear-to-nonlinear transition. 

\subsubsection{Velocity dispersion $\sigma_h^2(m_h,a)$ from linear theory}
\label{sec:sigma_h}
From linear perturbation theory, the variance of the large-scale peculiar velocity field smoothed on a comoving scale $R$ is \citep{STRAUSS:1995-The-density-and-peculiar-velocity-fields} 
\begin{equation}
\label{eq:sigma_h_full}
\sigma_h^2(R,a) = \frac{a^2H^2\left[f(\Omega_m)\right]^2}{6\pi^2} \int_0^\infty P(k,a)\,W^2(kR)\,dk, 
\end{equation}
where $f(\Omega_m)\approx\Omega_m^{0.55}$ is the linear growth-rate factor, $P(k,a)$ is the matter power spectrum, and $W(kR)$ is the real-space top-hat window function in Fourier space,
\begin{equation}
\label{eq:tophat}
W(kR) = \frac{3\left[\sin(kR) - kR\cos(kR)\right]}{(kR)^3}. 
\end{equation}
The halo mass enclosed within the comoving radius $R$ is
\begin{equation}
\label{eq:mass-radius}
m_h = \frac{4\pi}{3}\,\bar{\rho}_0\,R^3, \quad\Longrightarrow\quad R = \left(\frac{3\,m_h}{4\pi\,\bar{\rho}_0}\right)^{1/3},
\end{equation}
where $\bar{\rho}_0 = \Omega_m \rho_{c,0}$ is the comoving mean matter density.

Taylor-expanding Eq. \eqref{eq:tophat} for $kR \ll 1$ leads to 
\begin{equation}
\label{eq:W2-expand-exact} 
W^2(kR) = 1 - \frac{(kR)^2}{5} + \mathcal{O}\!\left[(kR)^4\right]. 
\end{equation}
Substituting into Eq. \eqref{eq:sigma_h_full} and defining the leading-order integrals
\begin{equation}
\label{eq:I0-def}
I_0(a) \equiv \int_0^\infty P(k,a)\,dk, \qquad I_2(a) \equiv \int_0^\infty k^2\,P(k,a)\,dk,
\end{equation}
so that
\begin{equation}
\label{eq:sigma_h_twoterms}
\sigma_h^2(R,a) = \frac{a^2H^2 f^2}{6\pi^2} \left[I_0(a) - \frac{R^2}{5}\,I_2(a)  + \mathcal{O}(R^4)\right].
\end{equation}
To evaluate $I_0$ in closed form, we approximate its dominant contribution near the matter-radiation equality wavenumber $k_{\rm eq}$ \citep{Aghanim:2021-Planck-2018-results--VI--Cosmo}:
\begin{equation}
\label{eq:I0-approx}
I_0(z=0) \approx P(k_{\rm eq},\,z{=}0)\,k_{\rm eq}.
\end{equation}
The equality wavenumber and its power spectrum are \citep{Aghanim:2021-Planck-2018-results--VI--Cosmo}
\begin{equation}
\label{eq:keq-exact}
k_{\rm eq}
= \frac{H_0}{c}\sqrt{\frac{2\Omega_m^2}{\Omega_{\rm rad}}}, \qquad P(k_{\rm eq},\,z{=}0) = \frac{2\pi^2\,A_s}{k_{\rm eq}^3\,a_{\rm eq}^2},
\end{equation}
where $A_s$ is the primordial scalar amplitude and $a_{\rm eq}$ is the scale factor at matter-radiation equality. So that
\begin{equation}
\label{eq:I0-final}
I_0(z{=}0) = \frac{\pi^2 A_s\,\Omega_{\rm rad}\,c^2} {\Omega_m^2\,H_0^2\,a_{\rm eq}^2}.
\end{equation}
The leading-order velocity variance at $z=0$ is therefore
\begin{equation}
\label{eq:sigma_h0-exact}
 \sigma_{h}^2(R\rightarrow0,z{=}0) = \frac{H_0^2\,f^2}{6\pi^2}\cdot I_0(z{=}0) = \frac{f^2\,A_s\,\Omega_{\rm rad}\,c^2} {6\,\Omega_m^2\,a_{\rm eq}^2}.
\end{equation}
This is independent of $m_h$ at leading order, confirming that $\sigma_h$ is insensitive to halo mass on scales $m_h<m_h^*$. 

The large-scale peculiar velocity variance and the smoothed density variance are intimately related through linear perturbation theory. Recall the density variance smoothed on scale $R$,
\begin{equation}
\label{eq:sigma_delta_def} 
\sigma_\delta^2(R,a) \equiv \frac{1}{2\pi^2}\int_0^\infty k^2\,P(k,a)\,W^2(kR)\,dk,
\end{equation}
and the bulk velocity variance (Eq. \eqref{eq:sigma_h_full}), which contains the same integrand but with two fewer powers of $k$. The ratio of the two integrals defines the scale-dependent factor $\beta_n$, yielding
\begin{equation}
\label{eq:spectrum-relation}
\begin{split}
&\sigma_h^2(R,a) = \tfrac{1}{3}\beta_n\,a^2H^2f^2\,\sigma_\delta^2(R,a)\,R^2, \\
&\beta_n \equiv\frac{\displaystyle\int_0^{\infty}P(k,a)\,W^2(kR)\,dk}
     {\displaystyle\int_0^{\infty}P(k,a)\,(kR)^2\,W^2(kR)\,dk}.
\end{split}
\end{equation}
For a locally power-law spectrum $P(k)\propto k^{n_\mathrm{eff}}$, Eq. \eqref{eq:spectrum-relation} gives 
\begin{equation}
\label{eq:betan}
\beta_n = \frac{4(1-n_\mathrm{eff})} {(2-n_\mathrm{eff})(5-n_\mathrm{eff})(1+n_\mathrm{eff})},
\end{equation}
which depends only on the local spectral index $n_\mathrm{eff}$ and is \emph{insensitive to the scale $R$} ($n_\mathrm{eff}$ varying slowly with $R$), for $n_\mathrm{eff}>-1$; as $n_\mathrm{eff}\to-1$, $\beta_n$ diverges as $(1+n_\mathrm{eff})^{-1}$ because the velocity integral is then dominated by the largest scales. This follows because the power spectrum is a regularly varying function: on any sufficiently narrow range of $k$, it admits the local representation $P(k,a)\approx C(a)k^{n_\mathrm{eff}}$ with a slowly varying amplitude $C(a)$, so $\beta_n$ is determined by the spectral shape, not the absolute scale.

\subsubsection{Quasi-universal spectrum near $m_h^*$}
The relation \eqref{eq:spectrum-relation} has qualitatively different behavior on either side of the characteristic scale $m_h^*$. On large, linear scales 
($m_h\gg m_h^*$), $\sigma_h^2(R,a)\propto\sigma_\delta^2(R,a)R^2$ as predicted by linear theory. On small, nonlinear scales ($m_h\ll m_h^*$), the bulk velocity variance saturates to a scale-independent value $\sigma_h^2(a)$ from Eq. \eqref{eq:sigma_h0-exact} (halos are fully virialized and their bulk motions decouple from the local density gradient). At the transition $m_h\sim m_h^*$, both regimes must match in value and in logarithmic slope:
\begin{equation}
\sigma_h^2(a) \sim \sigma_h^2(R,a) = \tfrac{1}{3}\beta_n a^2H^2f^2\,\sigma_\delta^2(R,a)\,R^2,
\end{equation}
For a local index $n_\mathrm{eff}>-1$, the linear branch scales as $\sigma_h^2(R)\propto\sigma_\delta^2R^2\propto R^{-(1+n_\mathrm{eff})}$, whereas the saturated branch has zero slope. Matching the slopes therefore requires $n_\mathrm{eff}\to-1$: the linear branch flattens onto the plateau only for
\begin{equation}
\sigma_\delta^2(R)\propto R^{-2}\propto m_h^{-2/3}\Longrightarrow P(k)\propto k^{-1} \text{ near } m_h\sim m_h^*,
\label{eq:universal-spectrum}
\end{equation}
independent of the initial spectrum. In this limit $\beta_n$ diverges as $(1+n_\mathrm{eff})^{-1}$ (Eq.~\eqref{eq:betan}), while $\sigma_\delta^2R^2$ becomes scale independent; their product stays finite for CDM spectrum at the saturated value, consistent with $\sigma_h$ being set by super-halo scales. This is the \emph{quasi-universal spectrum} generated near the linear-to-nonlinear transition, and it is a direct consequence of gravity. The emergence of $P(k)\propto k^{-1}$ near $m_h^*$ can be understood as the fixed point of nonlinear gravitational dynamics. Any initial spectrum is attracted toward it on scales where nonlinear collapse has fully processed the density field.
\begin{itemize}
\item \noindent \textbf{Gravitational fixed point.} The effective spectral index $n_\mathrm{eff}=-1$ is selected by the condition $\sigma_\delta^2(R)\propto R^{-2}$, which is precisely the inverse of the scale dependence of the gravitational potential energy $\Phi\propto Gm/R\propto \bar\rho R^2$ for halos at fixed overdensity, so that $\sigma_\delta^2\Phi$ is scale independent. Near $m_h^*$, the dynamics is governed by halo potentials rather than the primordial spectrum, so the density field self-organizes to the scale-invariant state consistent with the gravitational energy budget.

\item \noindent \textbf{Erasure of initial conditions.} For $n_\mathrm{eff}=-1$ the compensated spectrum $P(k)k$ is flat, meaning equal velocity variance per logarithmic scale interval. Each logarithmic scale interval contributes equally to the velocity variance (Eq. \eqref{eq:sigma_h_full}). Gravitational kinetic energy is distributed uniformly across scales, with no preferred scale. This is the condition for statistical equipartition across scales: repeated mergers and smooth accretion have erased memory of the primordial spectrum and driven the halo population toward the quasi-universal state. 

\item \noindent \textbf{Temporal evolution of the plateau.} As the universe expands and $m_h^*$ grows ($m_h^*\propto a^{3/2}$ in the nonlinear regime), the $k^{-1}$ plateau progressively extends to larger scales (smaller $k$), encroaching on the primordial-spectrum-dominated regime. This is directly visible in the compensated power spectrum $P(k)k$ of Fig. \ref{fig:3-1}, where the flat plateau broadens toward lower $k$ with decreasing redshift.
\end{itemize}

\subsubsection{Virial velocity dispersion $\sigma_v^2(m_h,a)$ from virial theorem}
\label{sec:sigma_v}
For a virialized halo of mass $m_h$ defined by a fixed overdensity $\Delta_{\rm vir}$ relative to the mean background density $\bar{\rho}(a) = \Omega_m\rho_{c}(a)$, the virial radius satisfies
\begin{equation}
\label{eq:rvir-exact}
m_h= \frac{4\pi}{3}\,\Delta_{\rm vir}\,\bar{\rho}(a)\,r_{\rm vir}^3\quad\Longrightarrow\quad r_{\rm vir}= \left(\frac{3\,m_h}{4\pi\,\Delta_{\rm vir}\,\bar{\rho}(a)}   \right)^{1/3}.
\end{equation}
The virial theorem for a self-gravitating system gives 
\begin{equation}
\label{eq:virial-exact}
\sigma_v^2(m_h,a)= \alpha_v\frac{\,G\,m_h}{r_{\rm vir}},
\end{equation}
where $\alpha_v$ is a dimensionless coefficient of order unity depending on the halo density profile ($\alpha_v = 1/2$ for a singular isothermal sphere; $\alpha_v \approx 0.45$ for an NFW profile with concentration $c_{\rm NFW}\approx 5$). Now apply the exact Friedmann equation,
\begin{equation}
\label{eq:Friedmann-exact}
\bar{\rho}(a)= \frac{3\,H^2(a)\,\Omega_m}{8\pi\,G},
\end{equation}
to eliminate $G\bar{\rho}$:
Substituting into Eq.~\eqref{eq:virial-exact}, we obtain
\begin{align}
\label{eq:sigma_v_step2}
\sigma_v^2(m_h,a) &= \frac{1}{2}\alpha_v\Delta_{vir}\Omega_mH^2r_{vir}^2.
\end{align}
A fixed $\Delta_{\rm vir}$ relative to $\bar\rho(a)$, and hence $\sigma_v^2\propto m_h^{2/3}a^{-1}$ at fixed mass, holds before the crossover $a_*$. After $a_*$, the scaling $\sigma_v^2\propto m_h^{2/3}$ at a given time still holds, but the dynamical density of halos near $m_h^*$ no longer tracks $\bar\rho(a)$ (Section~\ref{sec:3}).

\subsubsection{Crossing condition: exact derivation of $m_h^*$}
\label{sec:crossing-exact}
For the same mass scale $m_h$, the virial radius $r_{vir}$ and the comoving radius $R$ are related as $aR/r_{vir}=(\Delta_{vir})^{1/3}$. From Eqs. \eqref{eq:sigma_v_step2} and \eqref{eq:spectrum-relation}, the ratio between the two velocity dispersions is: 
\begin{equation}
\label{eq:crossing-explicit}
\frac{\sigma_{v}^2(m_h,a)}{\sigma_h^2(m_h,a)} = \frac{3\,\alpha_v\left(\Delta_{vir}\right)^{1/3}\Omega_m}{2\beta_nf(\Omega_m)^2\sigma^2_{\delta}} =\frac{\gamma_v}{\sigma^2_{\delta}(m_h,a)}.
\end{equation}
The ratio of virial to halo velocity dispersions reduces to $\sigma_{v}^2/\sigma_h^2 = \gamma_v/\sigma_\delta^2$, with $\gamma_v \sim 1$ absorbing the geometric and cosmological prefactors. Since the linear-to-nonlinear transition occurs at $\sigma_\delta^2(m_h^*,a) \sim 1$, the crossing condition $\sigma_v^2(m_h^*) = \sigma_h^2(m_h^*)$ naturally identifies $m_h^*$ as this transition scale. Because $\gamma_v$ is set by $\Delta_{\rm vir}^{1/3}$, $\beta_n$ and the profile coefficient $\alpha_v$, and because $m_h^*$ refers to virialized rather than Lagrangian scales, $m_h^*$ can exceed $M_{PS}^*$ by large factors at early times ($m_h^*/M_{PS}^*\sim50$ at $z\approx2$; Section~\ref{sec:3}). Before $a_*$, $\sigma_h^2\propto a$ and $\sigma_v^2\propto a^{-1}m_h^{2/3}$ give $m_h^*\propto a^3$; after $a_*$, both depart from these laws and $m_h^*\propto a^{3/2}$ (Fig.~\ref{fig:4}). The scale $m_h^*$, verifiable directly from $N$-body simulations via the crossing condition, carries a dual interpretation: it is simultaneously the characteristic mass-transport scale of the Fokker-Planck diffusion process and the dynamical scale at which inter-halo bulk kinetic energy equals intra-halo virial kinetic energy.

\added{
\section{Summary of simulation datasets}
\label{sec:simdesc}
For completeness, we briefly summarize here the $N$-body and hydrodynamical simulation datasets referenced in Figs. \ref{fig:3-1}, to \ref{fig:5}.
\begin{itemize}
\item \noindent \textit{Virgo SCDM} \citep{Jenkins:1998-Evolution-of-structure-in-cold}: a dark-matter-only $N$-body simulation of the Virgo Consortium in a standard (unnormalized) Cold Dark Matter cosmology, evolving $\sim 2\times 10^7$ particles in a periodic box.
\item \noindent \textit{Illustris-1-Dark} \citep{Vogelsberger:2014-Introducing-the-Illustris-Project}: the dark-matter-only counterpart of the Illustris project, a cosmological box of side $106.5\,\mathrm{Mpc}$ evolved with $1820^3$ dark matter particles in a $\Lambda$CDM cosmology.
\item \noindent \textit{IllustrisTNG} \citep{Springel:2018-First-results-from-the-IllustrisTNG}: the updated, magnetohydrodynamical Illustris simulation suite including galaxy formation physics; the dark matter power spectra shown in Fig.~\ref{fig:3-1} use the TNG100.
\item \noindent \textit{Ghalo} \citep{Stadel:2009-Quantifying-the-heart-of-darkness}: a high-resolution zoom-in $N$-body simulation of a single Milky Way-mass halo ($M_{200}\approx 8\times10^{11}M_\odot$) resolved with $\sim3\times10^8$ particles, used to probe the innermost resolved density slope.
\item \noindent \textit{Via Lactea} \citep{Diemand:2008-Clumps-and-streams-in-the-local}: a zoom-in simulation of a Milky Way-mass halo with $\sim2\times10^9$ particles within the virial radius.
\item \noindent \textit{Aquarius} \citep{Springel:2008-The-Aquarius-Project}: a suite of six zoom-in simulations of Milky Way-mass haloes at multiple resolution levels (up to $\sim4\times10^9$ particles).
\item \noindent \textit{Dubinski} \citep{Dubinski:1991-The-Structure-of-Cold-Dark}: an early high-resolution $N$-body simulation of individual cluster and galaxy-scale haloes, among the first to resolve the inner cuspy structure of CDM haloes.
\item \noindent \textit{FIRE:DMO} \citep{McKeown:2022-Amplified-J-factors-in-the-Galactic-Centre}: the dark matter only counterpart of the FIRE (Feedback In Realistic Environments) zoom-in simulations.
\end{itemize}}

\bibliographystyle{Papers}
\bibliography{Papers}

\label{lastpage}
\end{document}